# Overview to the Hard X-ray Modulation Telescope (*Insight*-HXMT) Satellite

ShuangNan Zhang[1,2*],TiPei Li[1,2,3], FangJun Lu[1], LiMing Song[1], YuPeng Xu[1], CongZhan Liu[1], Yong Chen[1], XueLei Cao[1], QingCui Bu[1], Ce Cai[1,2], Zhi Chang[1], Gang Chen[1], Li Chen[4], TianXiang Chen[1], Wei Chen[1], YiBao Chen[3], YuPeng Chen[1], Wei Cui[1,3], WeiWei Cui[1], JingKang Deng[3], YongWei Dong[1], YuanYuan Du[1], MinXue Fu[3], GuanHua Gao[1,2], He Gao[1,2], Min Gao[1], MingYu Ge[1], YuDong Gu[1], Ju Guan[1], Can Gungor[1], ChengCheng Guo[1,2], DaWei Han[1], Wei Hu[1], Yan Huang[1],Yue Huang[1,2], Jia Huo[1], ShuMei Jia[1], LuHua Jiang[1], WeiChun Jiang[1], Jing Jin[1], YongJie Jin[5], Lingda Kong[1,2], Bing Li[1], ChengKui Li[1], Gang Li[1], MaoShun Li[1], Wei Li[1], Xian Li[1], XiaoBo Li[1], XuFang Li[1], YanGuo Li[1], ZiJian Li[1,2], ZhengWei Li[1], XiaoHua Liang[1], JinYuan Liao[1], Baisheng Liu[1], GuoQing Liu[3], HongWei Liu[1], ShaoZhen Liu[1], XiaoJing Liu[1], Yuan Liu[6], YiNong Liu[5], Bo Lu[1], XueFeng Lu[1], Qi Luo[1,2], Tao Luo[1], Xiang Ma[1], Bin Meng[1], Yi Nang[1,2], JianYin Nie[1], Ge Ou[1], JinLu Qu[1], Na Sai[1,2], RenCheng Shang[3,], GuoHong Shen[7], XinYing Song[1], Liang Sun[1], Ying Tan[1], Lian Tao[1], WenHui Tao[1], YouLi Tuo[1,2], Chun-Qin Wang[7], GuoFeng Wang[1], HuanYu Wang[1], Juan Wang[1], WenShuai Wang[1], YuSa Wang[1], XiangYang Wen[1], BoBing Wu[1], Mei Wu[1], GuangCheng Xiao[1,2], Shuo Xiao[1,2], ShaoLin Xiong[1], He Xu[1], LinLi Yan[1,2], JiaWei Yang[1], Sheng Yang[1], YanJi Yang[1], Qibin Yi[1], JiaXi Yu[1], Bin Yuan[7], AiMei Zhang[1], ChunLei Zhang[1], ChengMo Zhang[1], Fan Zhang[1], HongMei Zhang[1], Juan Zhang[1], Liang Zhang[1], Qiang Zhang[1], ShenYi Zhang[7], Shu Zhang[1], Tong Zhang[1], Wei Zhang[1,2], WanChang Zhang[1], WenZhao Zhang[4], Yi Zhang[1], Yue Zhang[1,2], YiFei Zhang[1], YongJie Zhang[1], Zhao Zhang[3], Zhi Zhang[1], ZiLiang Zhang[1], HaiSheng Zhao[1], JianLing Zhao[1], XiaoFan Zhao[1,2], ShiJie Zheng[1], Yue Zhu[1], YuXuan Zhu[1], Renlin Zhuang[3], and ChangLin Zou[1] (The *Insight*-HXMT team)

[1] *Key Laboratory of Particle Astrophysics, Institute of High Energy Physics, Chinese Academy of Sciences, Beijing 100049, China*
[2] *University of Chinese Academy of Sciences, Chinese Academy of Sciences, Beijing 100049, China*
[3] *Department of Physics, Tsinghua University, Beijing 100084, China*
[4] *Department of Astronomy, Beijing Normal University, Beijing 100088, China*
*5 Department of Engineering Physics, Tsinghua University, Beijing 100084, China*
*6 National Observatories, Chinese Academy of Sciences, Beijing 100101, China*
*7 National Space Science Center, Chinese Academy of Sciences, Beijing 100190, China*

As China's first X-ray astronomical satellite, the Hard X-ray Modulation Telescope (HXMT), which was dubbed as *Insight*-HXMT after the launch on June 15, 2017, is a wide-band (1 – 250 keV) slat-collimator-based X-ray astronomy satellite with the capability of all-sky monitoring in 0.2-3 MeV. It was designed to perform pointing, scanning and gamma-ray burst (GRB) observations and, based on the Direct Demodulation Method (DDM), the image of the scanned sky region can be reconstructed. Here we give an overview of the mission and its progresses, including payload, core sciences, ground calibration/facility, ground segment, data archive, software, in-orbit performance, calibration, background model, observations and some preliminary results.

*Corresponding author (email: zhangsn@ihep.ac.cn)

# 1 Introduction

Since the launch of the first X-ray satellite, Uhuru [1], in 1970, soft X-ray astronomy satellites (covering <20 keV) have been extensively developed in the last century. The first hard X-ray telescope (covering >20 keV), HEAO-1 (A4) [2], was launched in 1977, which was not much later than Uhuru; however, the developments with regard to detectors and astrophysics in hard X-rays have been much less advanced than that in the soft X-ray band. High-energy photons in the hard X-ray band are difficult to collect via mirror reflection, and mature technology to collect these photons had not been developed until recently. Therefore, hard X-ray observations usually suffer from high background and low sensitivity. Historically, hard X-ray telescopes were mainly developed in three stages. First-generation instruments mainly used collimators to confine their field-of-views (FOVs), with a maximum typical effective area of up to 1000–2000 $cm^2$. Then, most hard X-ray telescopes, such as the telescope on board the European satellite INTEGRAL [3] and the burst alert telescope on board the US lead satellite *Swift*[4], employed ~2000 coded masks at an angular resolution of ~10 arcminutes to improve the imaging capability. These telescopes have effective areas of ~3000–5000 $cm^2$ and large FOVs of a few hundred to several thousand $deg^2$. Finally, to further improve the sensitivity and imaging capability, the first focusing hard X-ray telescope, NuSTAR[5], was launched in 2012; however, its effective area was reduced to <200 $cm^2$ above 20 keV, making it less effective in studying the timing behaviors of hard X-ray emissions.

Observational X-ray astronomy in China began in the late 1970s. Since then, a series of balloon-borne X-ray instruments have been developed for observing bright X-ray pulsars and X-ray binaries [6,7]. An innovative image reconstruction method, known as the direct demodulation method, was developed to obtain images using non-imaging instruments [8,9,10]. Based on the direct demodulation method and balloon observations, the Hard X-ray Modulation Telescope (HXMT) project was proposed [11,12].

The concept of *Insight*-HXMT was first proposed in 1993 and evolved into a mission covering spectra of 1–250 keV and higher. *Insight*-HXMT primarily aimed at conducting all-sky surveys within 20–250 keV to locate a large sample of obscured supermassive black holes (BHs) and to build an active galactic nuclei (AGN) catalog, which is essential for probing the extragalactic hard X-ray diffuse emission. However, this objective has already been achieved following the launch of the European INTEGRAL mission in 2002 and the US *Swift* mission in 2004 [13,14], well before the *Insight*-HXMT was officially funded in 2011. We thus added two sets of X-ray instruments covering 1–15- and 5–30-keV spectra to observe X-ray binary systems (XRBs) in their bright states or outbursts over a broad energy band. The orientations of the collimators are optimized for Galactic plane scanning survey and monitoring instead of all-sky surveys for AGNs. In the final stage of development before its launch, the all-sky monitor capability of *Insight*-HXMT was extended to 0.2–3 MeV to further enhance its scientific capability.

*Insight*-HXMT was successfully launched as China's first X-ray astronomy satellite on June 15, 2017 at the JiuQuan Satellite Launch Center in Northwest China and presently offers a seamless service. After the launch, HXMT was officially dubbed as "Hui-Yan" (Smart Eye), or *Insight*-HXMT in honor of the famous Chinese physicist, Prof. He ZeHui, the founder of high-energy astrophysics in China. Herein, we introduce the *Insight*-HXMT mission and its progresses, including payloads, core sciences, ground calibration/facility, ground segment, data archive, software, in-orbit performance, calibration, background model, observations, and some preliminary results.

## 2 The *Insight*-HXMT Mission

### 2.1 Overview of the Mission

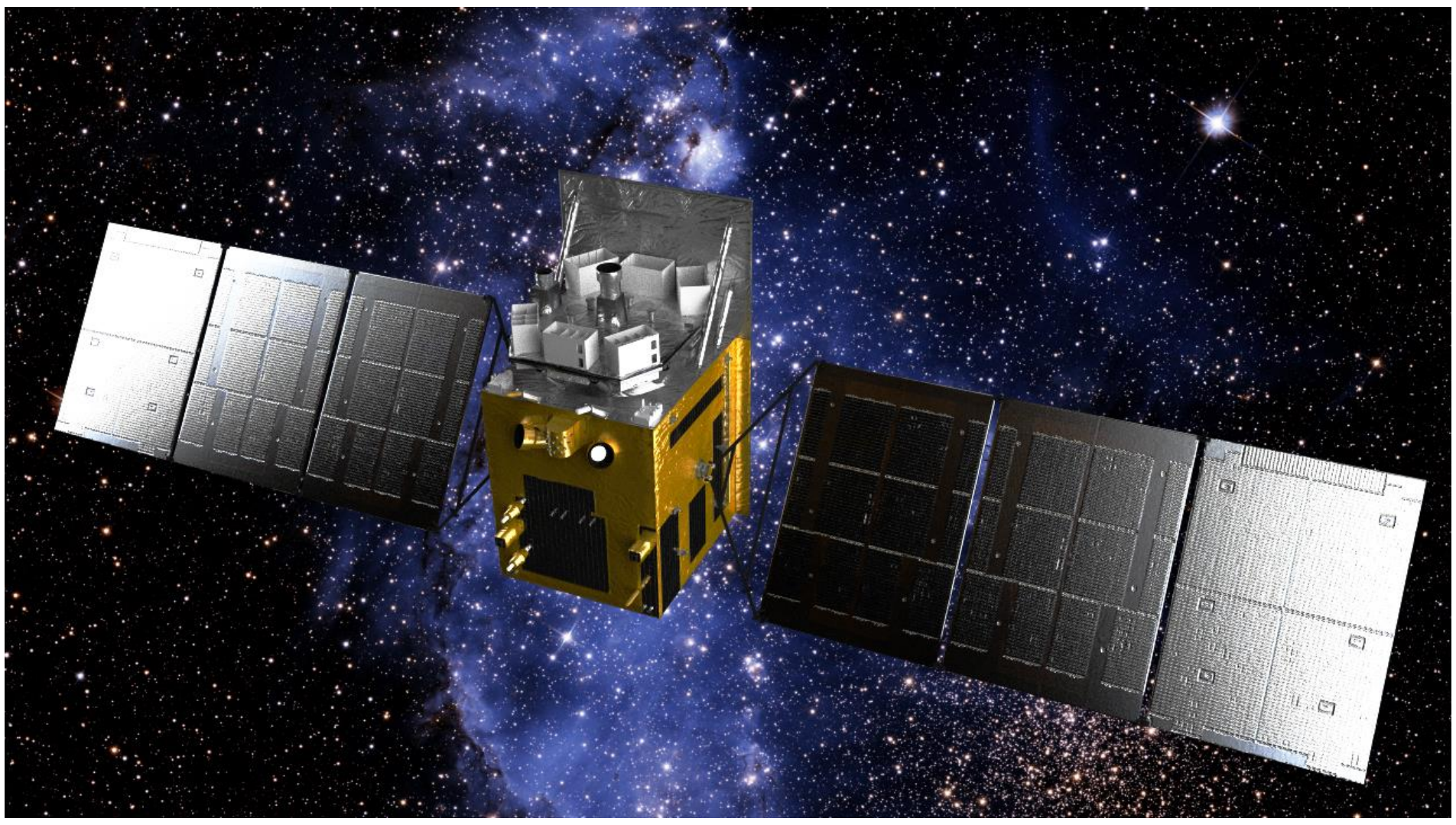

Figure 1 An artist's illustration of *Insight*-HXMT in space.

Figure 1 gives an artist's view of the *Insight*-HXMT satellite in space. It contains basically two cabins, the payload cabin and the platform cabin. The payload cabin is on the top of the satellite, in which all the telescopes are installed. The telescopes are all aligned to one direction so that they can observe the same source simultaneously. There is a sun-shielding board on one side of the satellite so that all the telescopes can avoid the direct irradiation of the sun, which will make the telescopes operate in a relatively low and stable temperature environment. The platform cabin is in the lower part of the satellite, which contains the service devices for attitude control, power supply, data management and down link, command receiving, and so on. The total weight of *Insight*-HXMT satellite is about 2500 kg.

*Insight*-HXMT runs in a low earth orbit with an altitude of 550 km and an inclination angle of 43°. The details of the mission profiles are listed in Table 1. There are three ground stations to receive the data from *Insight*-HXMT, which are in Hainan, Beijing, and Kashgar. To meet the requirements of the scientific observations, the satellite has three attitude control modes:

1) All sky survey mode: the sun-shielding board is perpendicular to the direction of the sun, and the satellite rotates slowly around the Sun direction so that the earth is out of the FOVs of the telescopes. In this mode, the whole sky can be covered in half a year.

2) Pointing observation mode: the satellite works in the 3-axis stabilized mode and the optical axis of the telescopes points to the observational target for a given duration.

3) Small region scan mode: the optical axis of the telescopes changes slowly in a planned track to cover a particular sky region. It is designed to find and locate new sources in a region, as well as monitor the flux variations of the known sources in the sky region.

Accordingly, *Insight*-HXMT has three observational modes: pointing, scan and Gamma-Ray Burst (GRB) modes. A pointing observation may have duration from one orbit (96 mins) up to 20 days, dedicated to spectral and timing studies. The Galactic plan can be surveyed with the small area scan mode. The entire Galactic plan is divided into 22 patches, each with a size of 20 deg×20 deg. The scan of one patch (see Figure 2) takes from 2 hours to about 5 days depending on combinations of the different scan parameters. The GRB mode was designed and implemented for the High Energy X-ray telescope (HE) shortly before launch. In this mode, the high voltage of the photo-multiplier tube (PMT) is reduced, so that the measured range of energy deposition in CsI can be increased to 0.2-3 MeV. This makes HE a unique high-energy gamma-ray all-sky monitor with a FOV of almost the entire sky, a large geometrical area (>1000 $cm^2$) and microsecond time resolution.

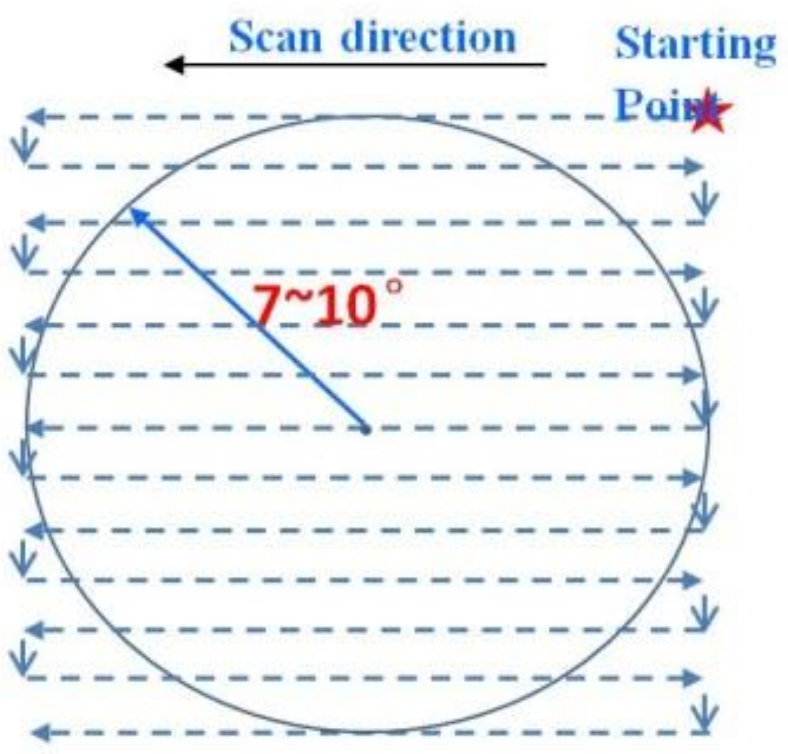

Figure 2 Demonstration of small sky area scan with *Insight*-HXMT.

As the extreme high particle flux within the South Atlantic Anomaly (SAA) not only causes large amount of event data which are useless for scientific observations, but also decreases life time of PMTs of HE significantly due to high anode current, we also introduced an SAA mode for the payloads. In this mode, the instruments do not record scientific events, and the high voltage of the PMTs are switched off to protect them. The SAA area is determined in two ways in parallel: 1) according to the measured total particle flux with the three Particle Monitors (PMs) of HE as well as proton and electron flux with the Space Environment Monitor (SEM). If the counting rate of any one of them is higher than the programmable threshold, the SAA mode will be triggered; and when all the counting rate keeps below the threshold at least 10 seconds, the SAA mode will be turned off automatically; 2) according to the predicted orbit time with the average particle flux determined with data recorded after launch. Each of the scenarios could be enabled or disabled with telecommands.

Table 1: Mission profiles of *Insight*-HXMT

| Items | Value |
|---|---|
| Orbit | 550 km, 43° |
| Weight | 2500 kg |
| Payload mass | 1000 kg |
| Lifetime | 4 years |
| Observation modes | With collimator: Pointed, Scanning<br>Without collimator: gamma-ray monitoring |
| Scanning speed | 0.01° /s, 0.03° /s, 0.06° /s |
| Pointing accuracy (3σ) | ±0.1° |
| Attitude measurement accuracy (3σ) | ±0.01° |
| Attitude stability | ±0.005° /S |
| Telemetry rate | 120 Mbps |
| ToO response time | ~5 hours |

## 2.2 Scientific Payload

As shown in Figure 3, there are three kinds of main scientific payloads onboard *Insight*-HXMT, which are the High Energy X-ray telescope (HE) [15], the Medium Energy X-ray telescope (ME) [16] and the Low Energy X-ray telescope (LE) [17]. To monitor the charged particle environment of the satellite, a supplementary instrument named Space Environment Monitor (SEM) is also installed on the satellite.

*Insight*-HXMT has a very wide energy band (1 keV-3 MeV) by using three different telescopes, each of which covers a different energy range. HE makes use of 18 NaI/CsI detectors covering the energy range of 20-250 keV for pointing and scan observation and the energy range of 0.2-3 MeV for all-sky gamma-ray monitoring. It has a total geometrical area of about 5100 $cm^2$. ME uses 1728 Si-PIN detectors with an energy range of 5-30 keV and a total geometrical area of 952 $cm^2$. LE uses swept charge device (SCD) as its detectors, which is sensitive in 1-15 keV with a total geometrical area of 384 cm$^2$.

All three telescopes use slat collimators to restrict their FOVs; the FOVs (FWHM) are 1.1°×5.7° and 5.7°×5.7° for HE, 1°×4° and 4°×4° for ME and 1°×6°, 6°×6° and 2.5°×60° for LE. There are also several detectors with blocked FOV for each telescope to measure the local background. Please refer to Figure 4 and Table 2 for details.

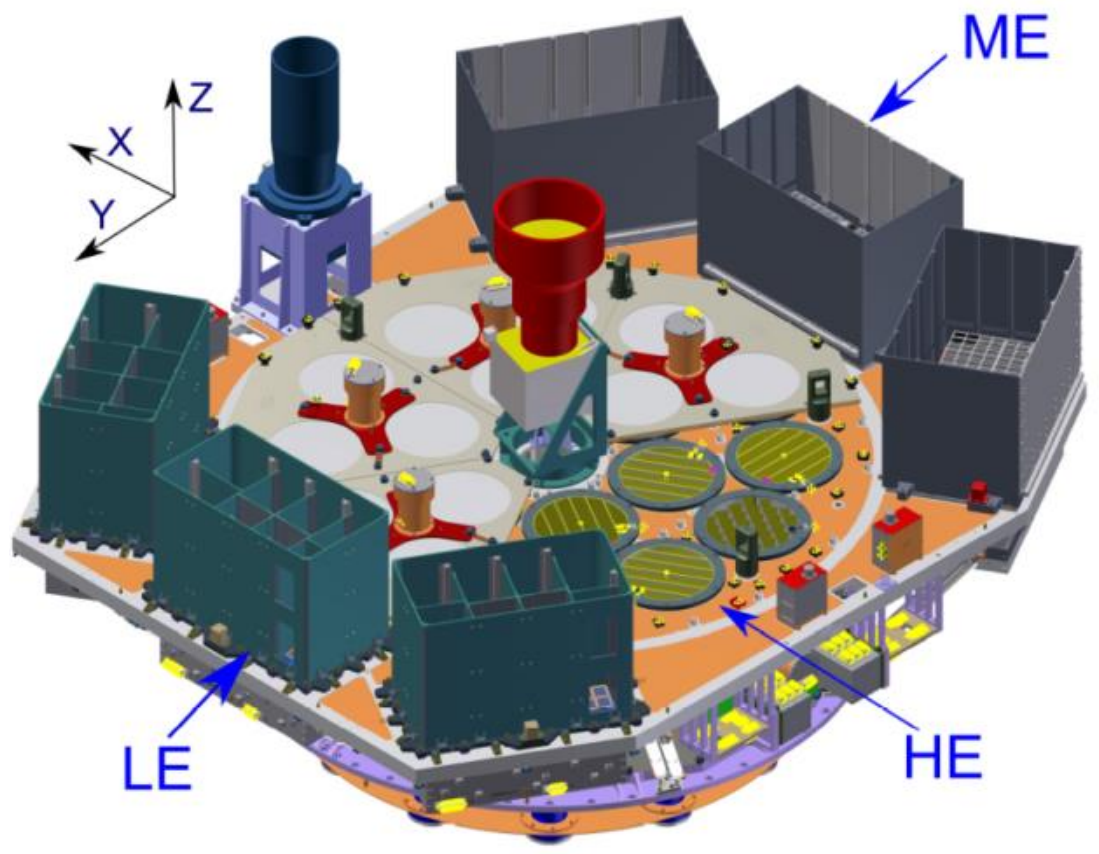

Figure 3 FOVs and their orientations for all telescopes.

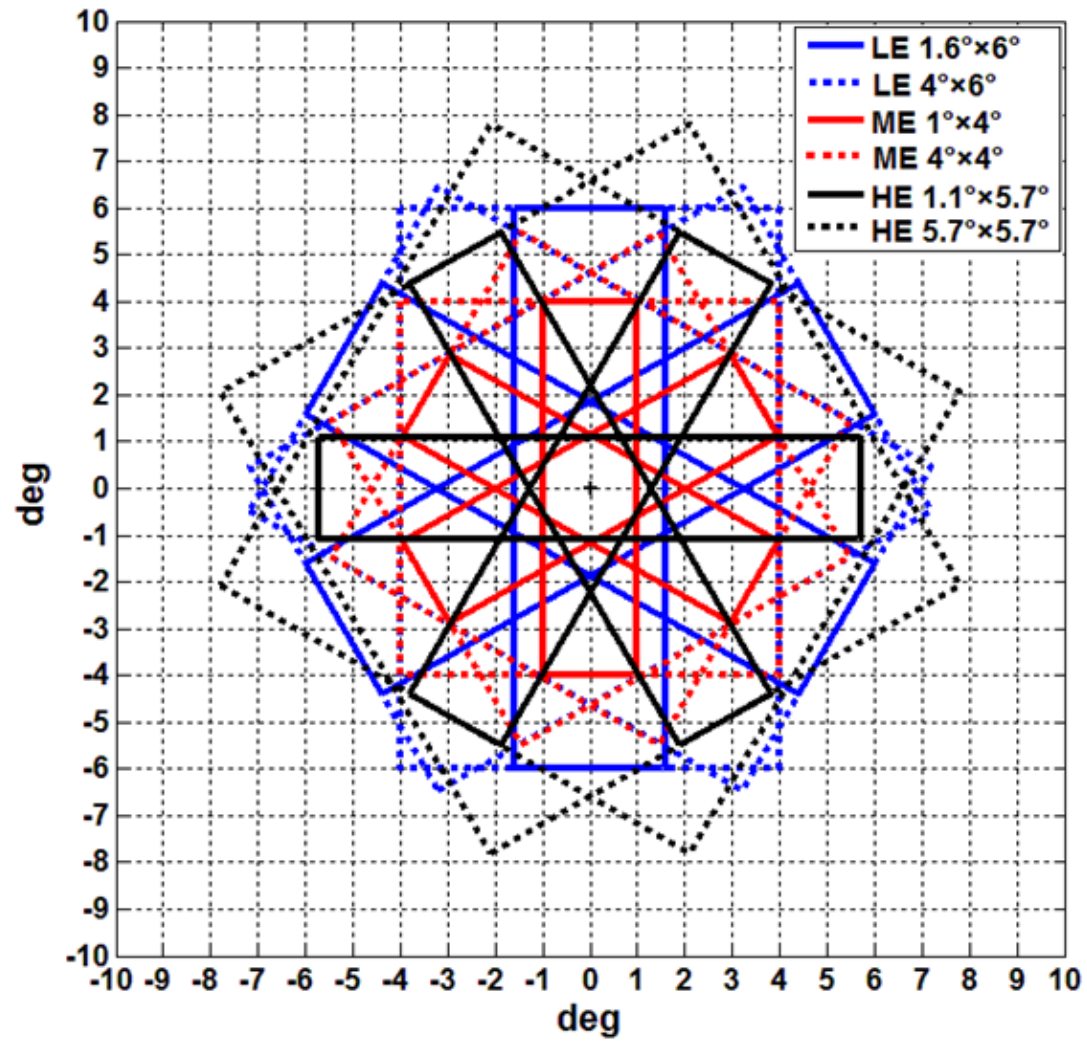

Figure 4 The FOVs of *Insight*-HXMT.

Table 2 FOVs and their orientations of all telescopes.

| Telescope | Number of collimators | Collimator FOV (FWHM) | Orientation in the payload coordinate (deg) |
|---|---|---|---|
| HE | 5 | 1.1°×5.7° | 0° |
| | 5 | 1.1°×5.7° | 60° |
| | 5 | 1.1°×5.7° | -60° |
| | 1 | 5.7°×5.7° | -60° |
| | 1 | 5.7°×5.7° | 60° |
| | 1 | 1.1°×5.7°（blocked） | 0° |
| ME | 15 | 1°×4° | 90° |
| | 15 | 1°×4° | 30° |
| | 15 | 1°×4° | -30° |
| | 2 | 4°×4° | 90° |
| | 2 | 4°×4° | 30° |
| | 2 | 4°×4° | -30° |
| | 1 | 1°×4°（blocked） | 90° |
| | 1 | 1°×4°（blocked） | 30° |
| | 1 | 1°×4°（blocked） | -30° |
| LE | 20 | 1.6°×6° | 90° |
| | 20 | 1.6°×6° | 30° |
| | 20 | 1.6°×6° | -30° |
| | 6 | 4°×6° | 90° |
| | 6 | 4°×6° | 30° |
| | 6 | 4°×6° | -30° |
| | 2 | 1.6°×6°，4°×6°（blocked） | 90° |
| | 2 | 1.6°×6°，4°×6°（blocked） | 30° |
| | 2 | 1.6°×6°，4°×6°（blocked） | -30° |
| | 2 | 60°×2.5° | 90° |
| | 2 | 60°×2.5° | 30° |
| | 2 | 60°×2.5° | -30° |

### 2.2.1 The High Energy X-ray telescope (HE)

HE consists of 18 NaI(Tl)/CsI(Na) phoswich detectors with a total geometrical area of about 5100 $cm^2$. HE can observe sources within the Field of View (FOV) defined by collimators in about 20-250 keV using mainly the NaI(Tl) scintillator either by pointing observation or by scan survey, and simultaneously monitor the whole sky in about 0.2-3 MeV using mainly the CsI(Na). The geometrical area of each detector is 283.5 $cm^2$, and its collimator defines FOVs of 1.1°×5.7° for 15 modules, 5.7°×5.7° for 2 modules, and one blind for measuring the local background. The orientations of the 18 FOVs are different, with a step size of 60 deg, as shown in Figure 4 and Table 2.

Each detector unit is a cylindrical NaI(Tl)/CsI(Na) phoswich scintillation detector with a diameter of 19 cm. The thickness of the NaI, the main detector, is 3.5 mm, while that of the shielding CsI(Na) is 40 mm. A 5-inch PMT is used to collect the fluorescence of both NaI and CsI (see Figure 5).

The tubes and frames of each collimator are cut from one aluminum cylinder. The thickness of the frames is 1 mm. On the frames a number of troughs are cut, and thin (0.15 mm) tantalum plates are inserted in these troughs. On the walls of the tubes and frames, thin (0.15 mm) tantalum strips are pasted. To further shield the background, a thick (2 mm) tantalum ring is installed surrounding each of the NaI/CsI detectors.

As well as using CsI(Na) in the phoswich mode as active shielding detector, *Insight*-HXMT is also equipped with a veto system covering the collimators, in order to detect and reject the charged particle events. The veto system consists of 18 plastic scintillators, with six of them on the top and 12 of them surrounding the HE main detector assembly.

The CsI detector is practically also an all-sky monitor beyond 200 keV: in NOrmal Mode (NOM) up to 600 keV, in Low

Gain Mode (LGM) or GRB mode [18] up to about 3 MeV (see Table 3).

Table 3: The energy coverage of *Insight*-HXMT in observing GRB.

| | Normal Mode | Low gain/GRB mode |
|---|---|---|
| NaI measured energy range | 20-250 keV | 100-1250 keV |
| CsI measured energy range | 40-600 keV | 200 – 3000 keV |

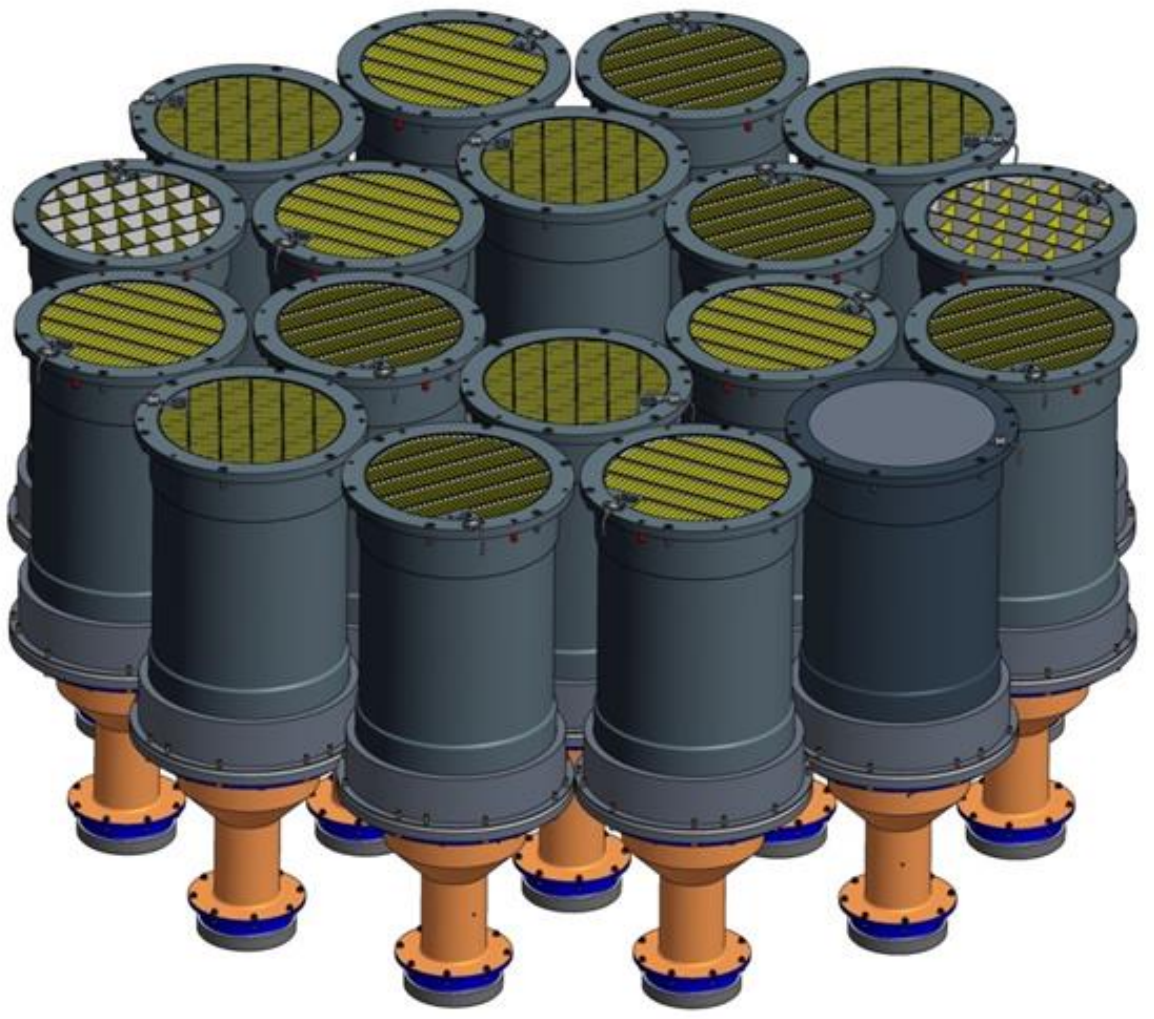

Figure 5 The detector modules and PMTs of *Insight*-HXMT/HE.

### 2.2.2 The Medium Energy X-ray telescope (ME)

ME is composed of three individual Si-PIN detector boxes working in 5-30 keV with a total geometrical area of 952 cm$^2$. In each box there are three units, and every unit contains six modules with 32 Si-PIN detector pixels each. The 32 Si-PIN detector pixels are read out by one VA32TA2 chip and each module works almost independently. There are 1728 Si-Pin detector pixels in total. Such a modularized design improves the overall reliability and makes the detector installation easier. Figure 6 shows the architecture of one ME detector box, and Figure 7 shows the detector array of the qualification model.

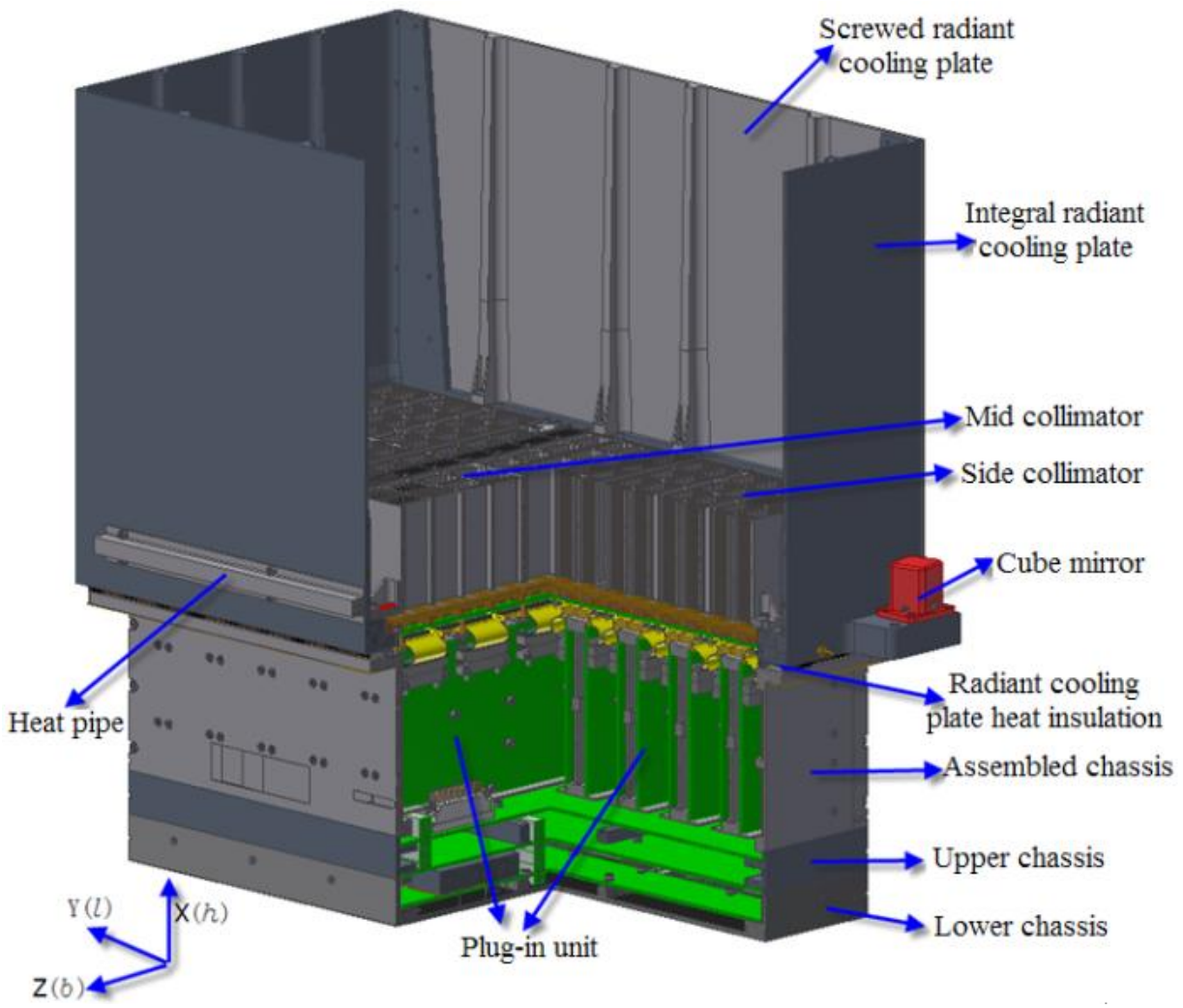

Figure 6 The architecture of one ME detector box.

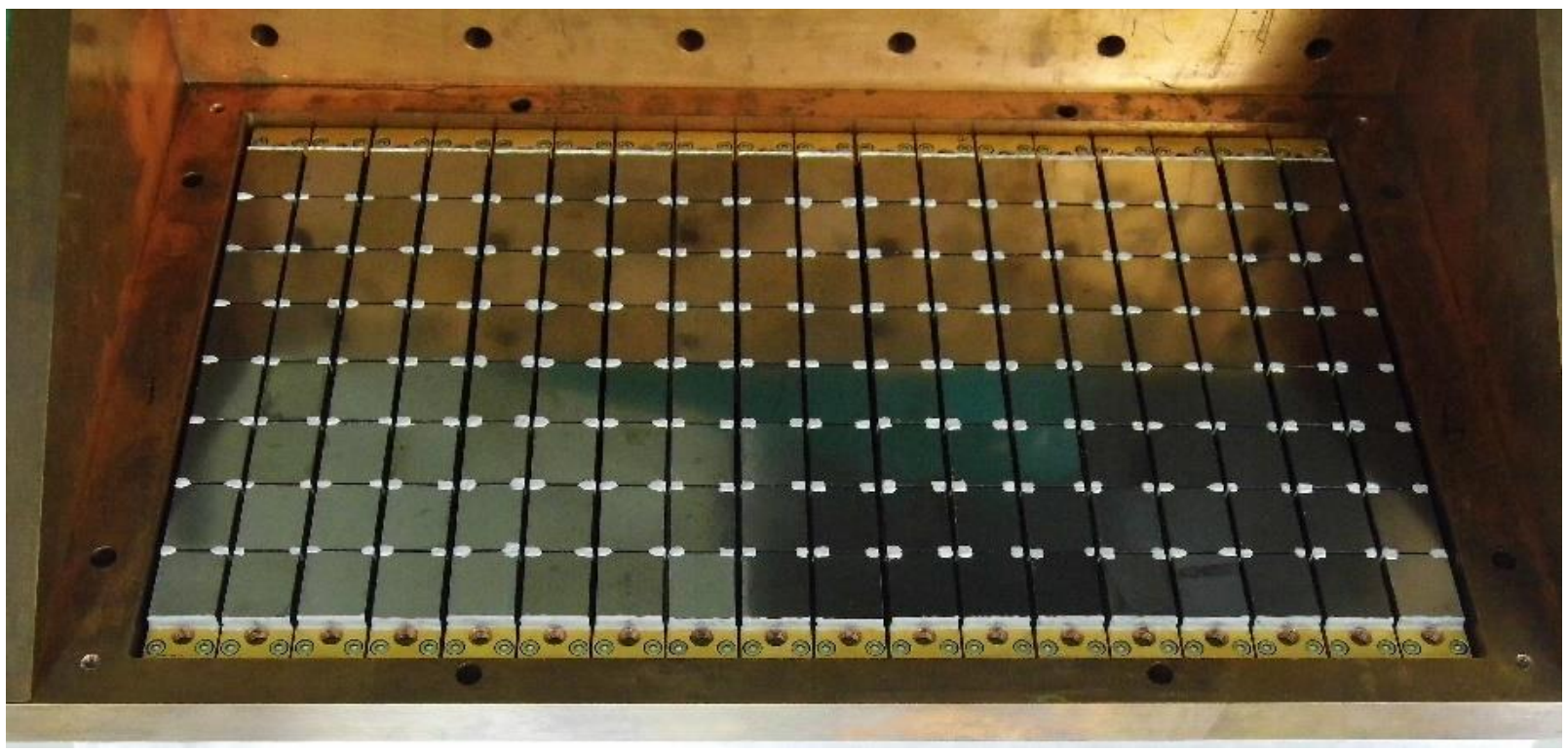

Figure 7 ME detector array of the flight model in calibration.

#### 2.2.3 The Low Energy X-ray telescope (LE)

LE covers the energy band of 1.0-15 keV. It uses SCD as the detector to achieve a high time resolution and a good energy resolution simultaneously. LE contains three detector boxes with a sun buffer each. The buffer could also be used as the radiator to cool the detectors. One LE detector box contains two modules and each module contains 16 SCD chips (CCD236), and every four CCD236 detectors share one collimator (see Figure 8). The total detection area of each module is 64 cm$^2$.

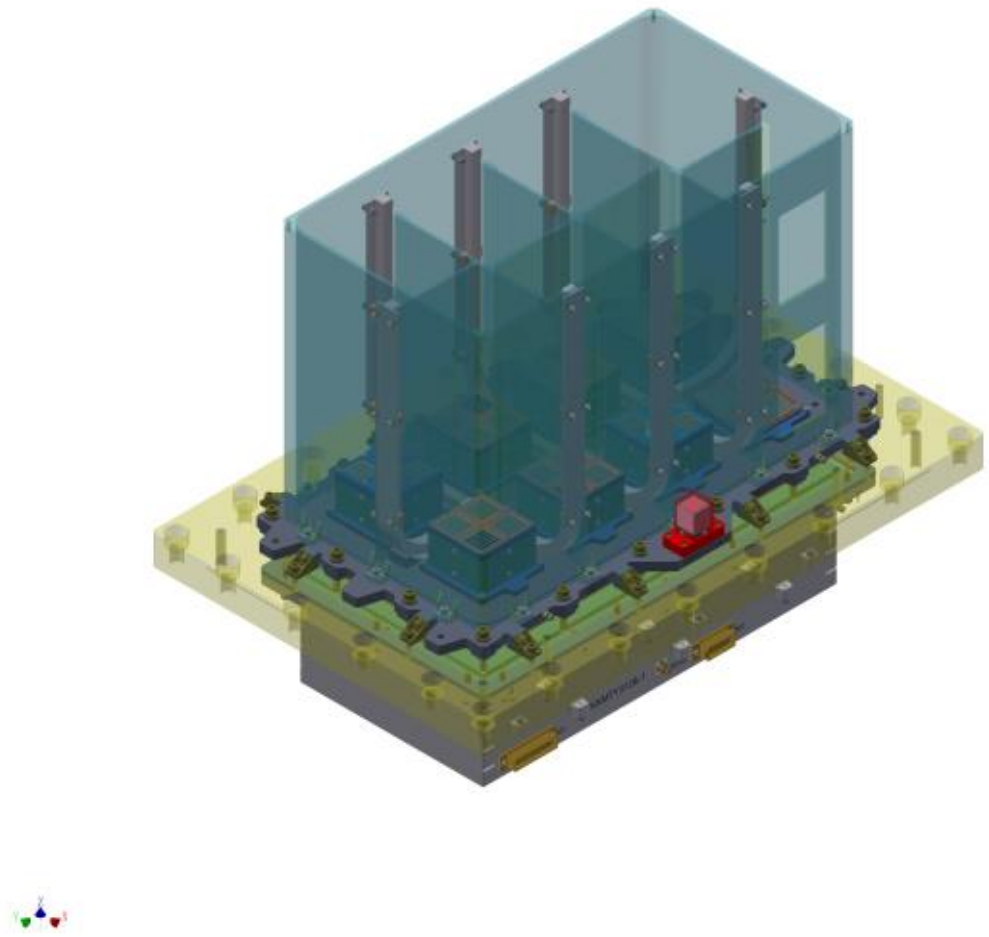

Figure 8 The structure of one LE detector box.

#### 2.2.4 The Space Environment Monitor (SEM)

The Space Environment Monitor (SEM; Figure 9) is to monitor the charged particle environment of the satellite, which can be used in the estimation of the in-orbit background of the telescopes. SEM has 18 individual detector units, among which one is to measure the spectrum of electrons in 0.4 to 1.5 MeV, one to measure the spectrum of protons in 3 to 150 MeV, and the other 16 units to measure the particle fluxes in different directions. Both the electron spectrometer and the proton spectrometer have seven energy channels. SEM gives one data point every second.

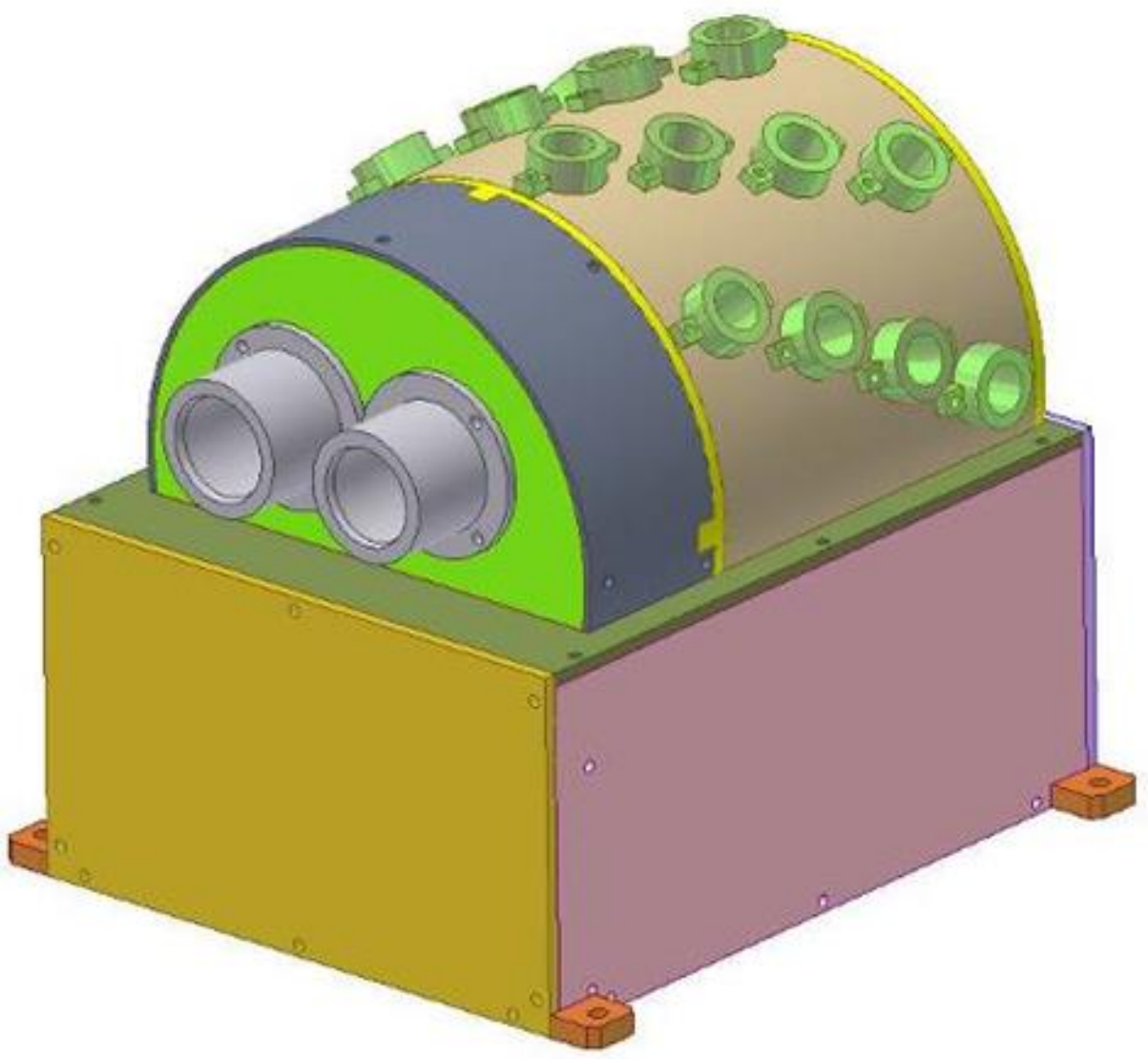

Figure 9 The configuration of SEM. The two detectors in grey color are the electron spectrometer (left) and the proton spectrometer (right), and all others measure particle fluxes from different directions.

### 2.2.5 Characteristics of HXMT

The key parameters of the mission are listed in Table 4. The effective areas of the three telescopes as functions of energy are shown in Figure 10.

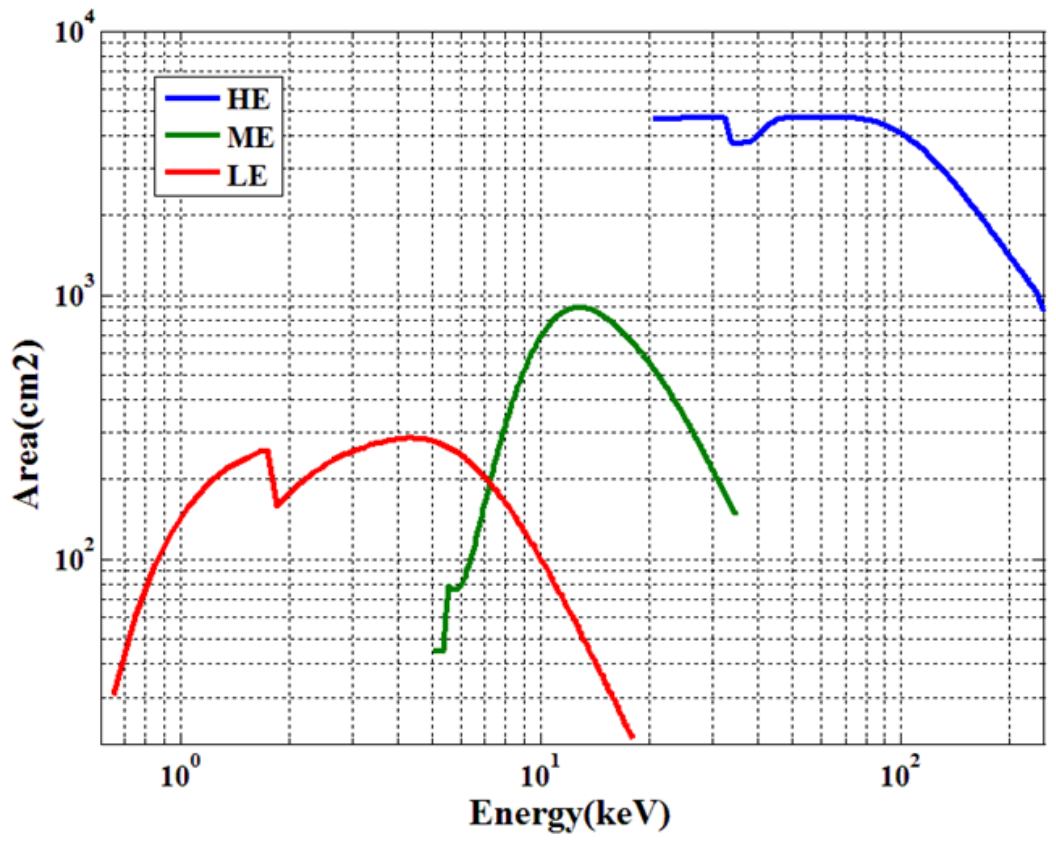

Figure 10 The effective area of *Insight*-HXMT.

Table 4 The main characteristics of the *Insight*-HXMT payloads.

| | |
|---|---|
| Detectors | LE: SCD, 384 $cm^2$<br>ME : Si-PIN, 952 $cm^2$<br>HE : NaI/CsI, 5000 $cm^2$ |
| Energy Range | LE: 1-15 keV<br>ME: 5-30 keV<br>HE: 20-250 keV |
| Time Resolution | HE: 25 μ s<br>ME: 280 μ s<br>LE: 1 ms |
| Energy Resolution | LE: 2.5% @ 6 keV<br>ME: 14% @ 20 keV<br>HE: 19% @ 60 keV |
| Data Handling Capa- | LE: ≤3 Mbps |
| bility | ME: ≤3 Mbps<br>HE: ≤300 kbps |

## 2.3 The Direct Demodulation Method

The DDM was proposed by Li and Wu in 1991, and has been successfully applied to analyze the observational data of different types of telescopes [19,20]. The basic point of this algorithm is to solve the observational equations that are largely distorted by statistical fluctuations. During the procedure of pursuing the proper solutions to this equation array, by introducing the physical constraints (for example the flux could not be negative) the distortions are largely suppressed and therefore one gets the desired solutions for both the flux and the position of the celestial object observed [9]. The *Insight*-HXMT Galactic plane survey map can be produced with this method and by properly handling/introducing background models as part of the response matrix in the iteration procedure [21]: either the particle background or the known sources can be encoded into the transformed response matrix as background models and tagged with scaling factors estimated or fixed in the iteration procedure as ‘pixels’ additional to the inferred sky region.

## 2.4 The *Insight*-HXMT ground calibration

Reliable and accurate calibrations are the basis for a space telescope to achieve its scientific goals. For *Insight*-HXMT, the responses of the detector depend strongly on the working conditions. As exampled in Figure 11, both the energy resolution and the signal amplitude recorded by the LE detector are sensitive to the operating temperatures; the similar effects also exist for the HE and ME detectors. The goal of the calibration is to measure/estimate the response matrices of the telescope at different working conditions.

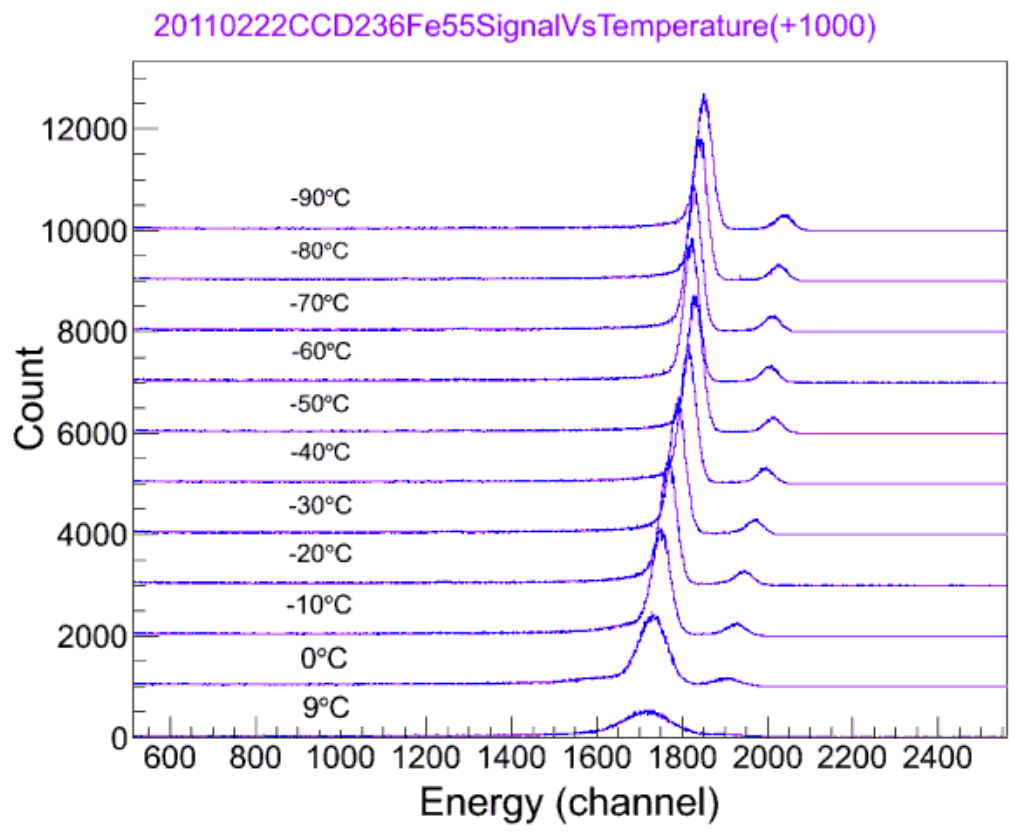

Figure 11 $^{55}$Fe spectra measured by the SCD detectors at different temperatures

There are two calibration facilities built specially for the *Insight*-HXMT [22,23]. The facility for HE works in 15-100 keV, and the one for ME and LE works in 0.8-30 keV. For both facilities the mono-energy X-ray beams are produced from the double crystal monochromators, with an intrinsic energy dispersion of roughly 0.1%-1%. Figure 12 shows the two facilities: the one for HE works in the air and under the room temperature, while the one for ME/LE is rather complex since supplementary devices like the vacuum chamber and cooling system have to be constructed to satisfy the working conditions of ME/LE.

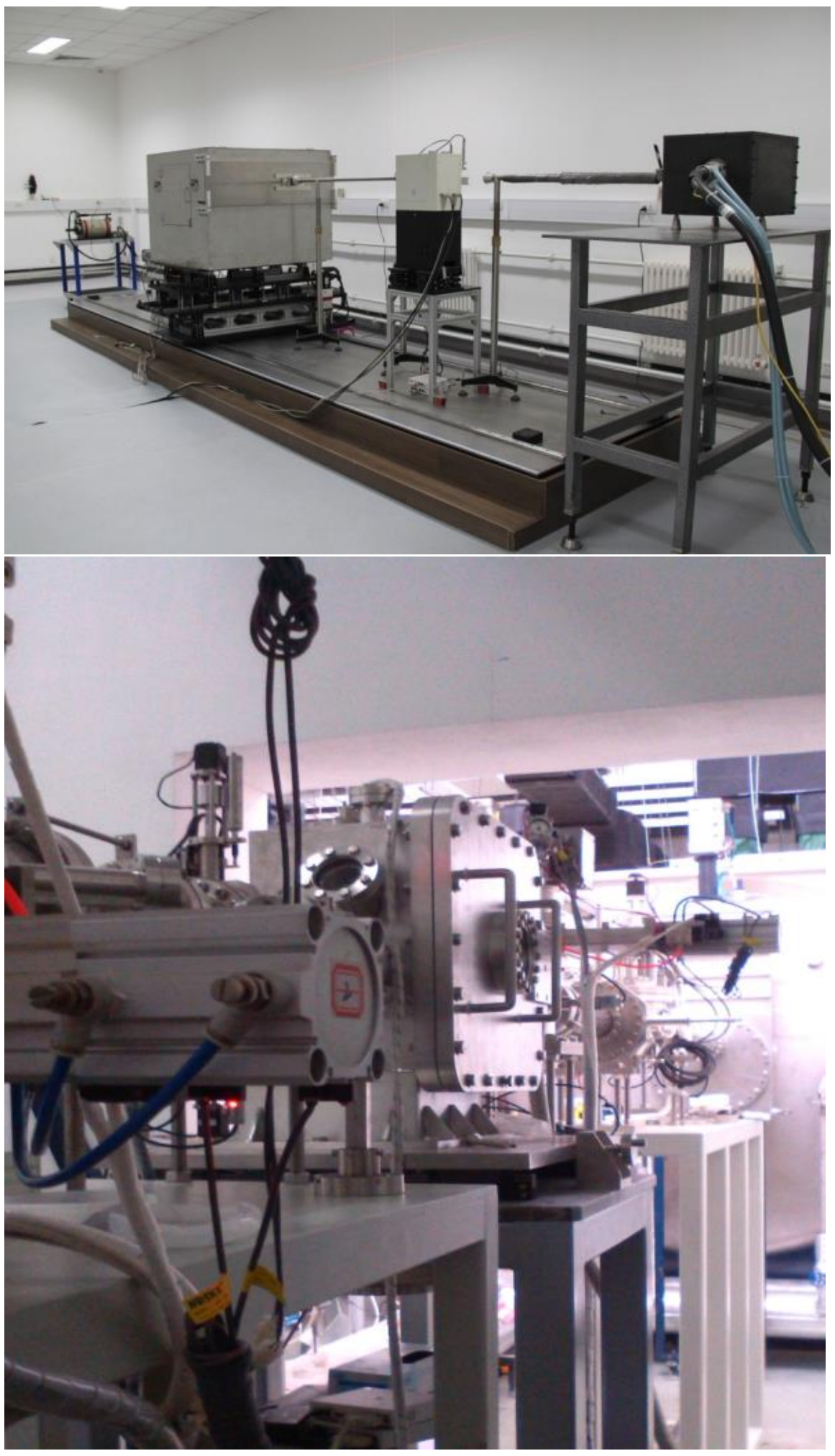

Figure 12 Calibration facilities for HE (upper), and for ME/LE (lower).

In addition to the above two facilities, there are two other vacuum chambers for the calibrations of ME and LE respectively. In the chambers, X-ray tubes are used to illuminate metal targets and hence produce florescent lines of iron, copper, molybdenum, and tin. These florescent lines are used to calibrate the variation of the energy response matrices of the detectors at different operating temperatures, in supplement to the monochromatic X-ray beams.

## 3. The core sciences

*Insight*-HXMT will carry out sensitive Galactic plane surveys in high cadence and is expected to build a Galactic source catalogue with sources showing up their behavior at short time scales at hard X-rays. *Insight*-HXMT will help to understand the low/hard state of XRB outbursts, thanks to the large geometrical area in 20-250 keV. At soft X-rays, *Insight*-HXMT can reach energies as low as 1 keV and, with the fast-readout capability of the non-imaging CCDs, *Insight*-HXMT/LE can ease the problems of photon saturation and pile-up effects in observing strong sources. Therefore, the source pointed observation of *Insight*-HXMT can provide broad band data with good statistics for sources that are bright persistently or undergoing an outburst. In addition, *Insight*-HXMT/HE serves as an excellent wide-field monitor for high-energy transient sources, i.e. GRBs, with the thick CsI detectors originally designed only for active shielding. Therefore, the core sciences of the *Insight*-HXMT are:

(1) to scan the Galactic plane and find new transient sources and to monitor the known variable sources;

(2) to observe X-ray binaries in broad energy band and study the dynamics and emission mechanism in strong gravitational or magnetic fields;

(3) to observe GRBs in a relatively rarely explored band from a few hundred keV to a few MeV.

## 4 The *Insight*-HXMT ground segment

The *Insight*-HXMT ground segment (HGS) consists of two main parts. One is the mission operation ground segment (MOGS), built and operated by the National Space Science Center (NSSC) of the Chinese Academy of Sciences (CAS), and the other is the science ground segment (SGS) [24] built and operated by the Institute of High Energy Physics (IHEP) of CAS. The functions of MOGS include operation commands, data receiving, data pre-processing, and permanent data archive after the mission is finished. The SGS has four centers responsible for users (HSUC), supports (HSSC), operation (HSOC) and data (HSDC), respectively. See Figure 13 for the details of the HGS. In the following we will introduce briefly the functions of SGS.

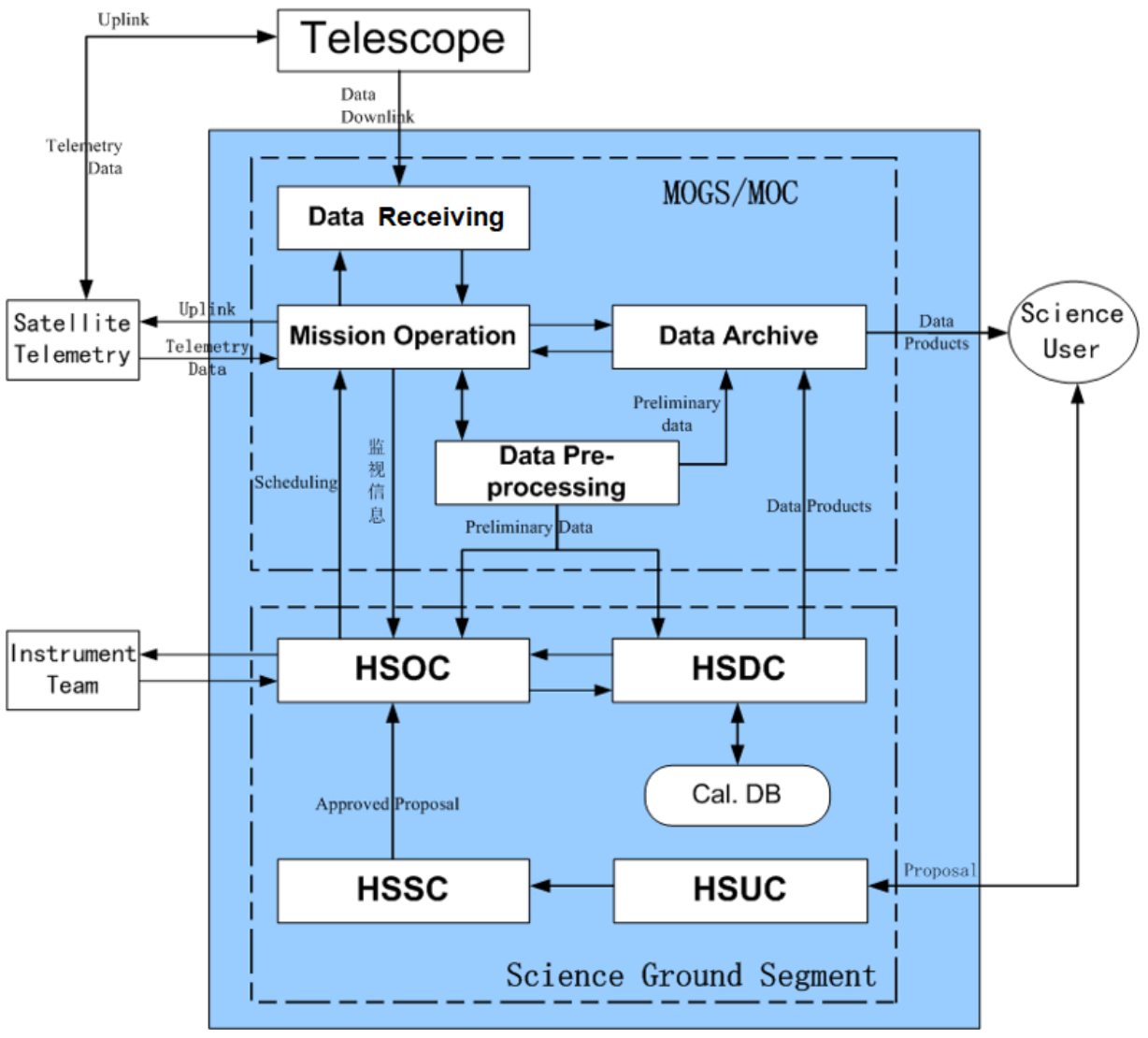

Figure 13 Overview of the *Insight*-HXMT HGS.

### 4.1 Data archive

All the *Insight*-HXMT data to be provided to users are recorded in the standard FITS format. The standard scientific data products have three levels: the level 0 data are the primary observational data, level 1 data are released to users, and level 2 are the scientific results obtained from the data. The HE data are classified into event, HE status, HE temperature, HE high voltage, count rate, dead time and data from the particle monitor. ME and LE data are made of event, status of the telescope, temperature, circuit parameters and the count rate. For more details see http://hsuc.ihep.ac.cn/FAQ/youknow.jsp and http://hsuc.ihep.ac.cn/web.

### 4.2 Analysis software

The *Insight*-HXMT Data Analysis Software (HXMTDAS) is developed in the software environment of HEAsoft, and is specified for analyzing the *Insight*-HXMT data from pointed observations. As shown in Figure 14, the input of HXMTDAS is the level 1 *Insight*-HXMT data products, and outputs are the cleaned and calibrated event files and high-level scientific products of e.g. light curves and energy spectra. HXMTDAS is run in three steps: (1) calibrate the event; (2) screen the event files and select the good time intervals with standard event selection criteria; (3) extract the scientific data products ready for timing and spectral analysis with tools available in HEAsoft. The HXMTDAS with its guide for installation is available at http://www.hxmt.org/index.php/enhome/analysis.

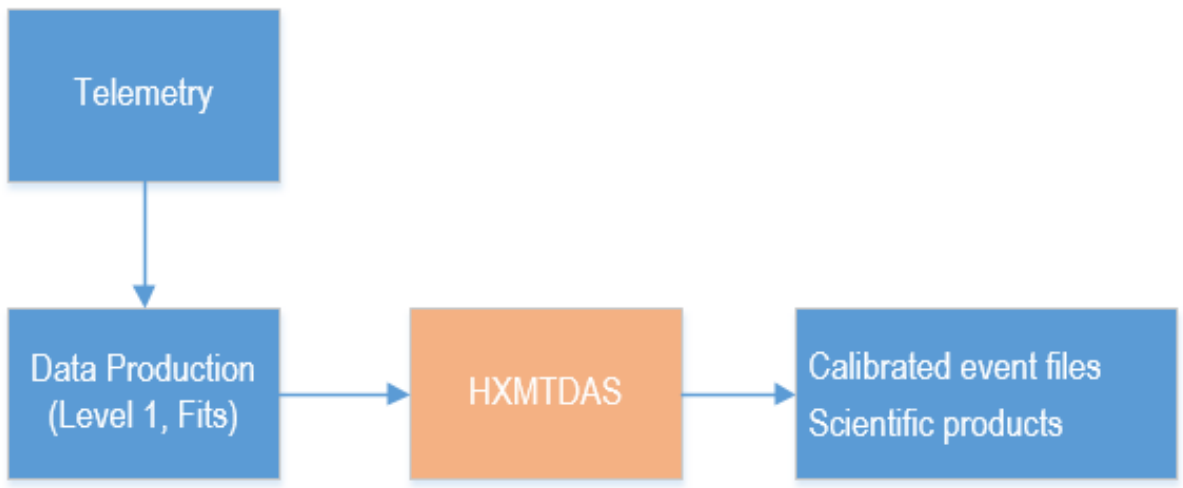

Figure 14 Input/output for HXMTDAS.

### 4.3 AO-1 and core science team

During 2016 August and September, the first round of the Announcement of Opportunity (AO) was released to the domestic astronomers calling for submissions of the *Insight*-HXMT core science proposals. After preliminary technical evaluation, 90 proposals were sent to the *Insight*-HXMT scientific committee for detailed review. These proposals request observations of over than 300 targets, which sum up to exposures of 331 days for normal observations and 1140 days for ToO observations. As shown in Figure 15, half of the requested exposure will be spent on the high cadence and high statistics observations of BH and NS XRBs. The *Insight*-HXMT core science team was formed with its members mainly from the *Insight*-HXMT project and the PIs/Co-Is of these core science proposals. The core science team is divided into eight groups to address the core sciences in the following aspects:

Group 1: Accretion X-ray binaries
Group 2: Galactic plane survey and diffuse emission
Group 3: Multi-wavelength observations
Group 4: Calibration and background model
Group 5: Pulsar navigation (autonomous navigation using X-ray observations of millisecond pulsars)
Group 6: Extragalactic sources
Group 7: Gamma-ray bursts and gravitational wave EM counterparts
Group 8: Non-accreting pulsars

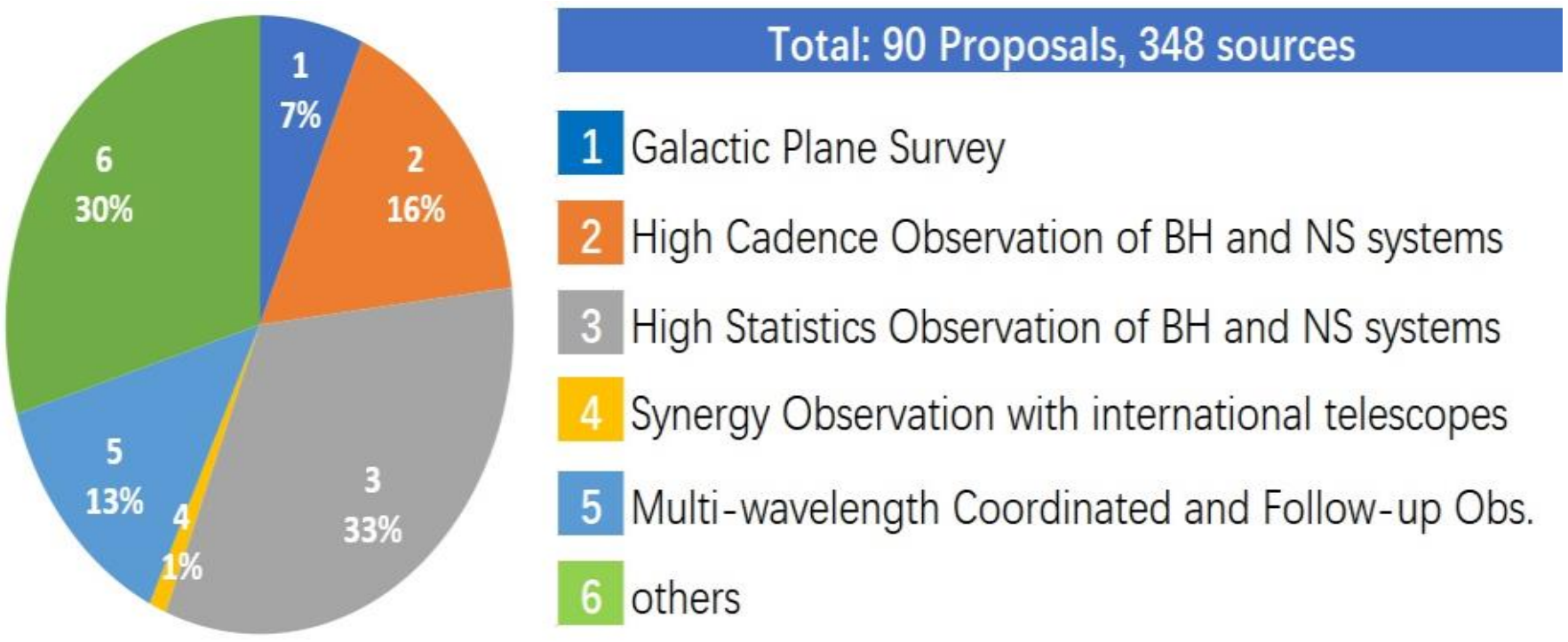

Figure 15 Research fields covered by the core science proposals.

### 4.4 Data policy

There are two major research programs for *Insight*-HMXT in-orbit observation, data analysis and related research. (1) *Insight*-HXMT Core Science Program. The observational data for core science program are shared within the core science team within the data proprietary period, which is nominally one year after the data products are available. (2) *Insight*-HXMT Guest Science Program. *Insight*-HXMT Guest Science Program is formulated by guest observers from both domestic and international institutions, where users can propose observations not already included in the Core Science Program. The approved proposals are scheduled by *Insight*-HXMT SGS, while the data are exclusively used only for guest science group within the data proprietary period, which is nominally one year after the data products are available. The above data policy is applicable to regularly scheduled observations and the first year Target of Opportunity (ToO) observations. From the second year, data from most of the *Insight*-HMXT ToO observations will be available immediately to facilitate joint multiwavelength and multi-mission studies. Further details of the *Insight*-HMXT data policy can be found on the *Insight*-HMXT website (hxmt.org).

### 4.5 **Observational plan and constraints**

*Insight*-HXMT SGS prepares the observation plans in term of the long, medium and short periods according to the corresponding science programs and the prediction of the orbit. These plans are made for observational periods of 1 year, 4 weeks and 2 days, with accuracy of up to 2 weeks, 2 days and 1 second, respectively. Making an observational plan is subject to the following constraints: 1) thermal control: the solar avoidance angle > 70°, and the angle between the direction of the sun and the X-Z plane of the payload < 10° (see Figure 16); 2) earth occultation: lasts about 30 minutes for almost every orbit; 3) South Atlantic Anomaly (SAA): roughly 8~9 orbits per day, with each lasting for about 15 minutes; 4) moon avoidance angle: > 6°.

The *Insight*-HXMT observational plans are available at http://www.hxmt.org/index.php/enhome/schedule. Please note that the *Insight*-HXMT short term plans are subject to frequent modifications due to the capability of *Insight*-HXMT's fast response to ToO observations.

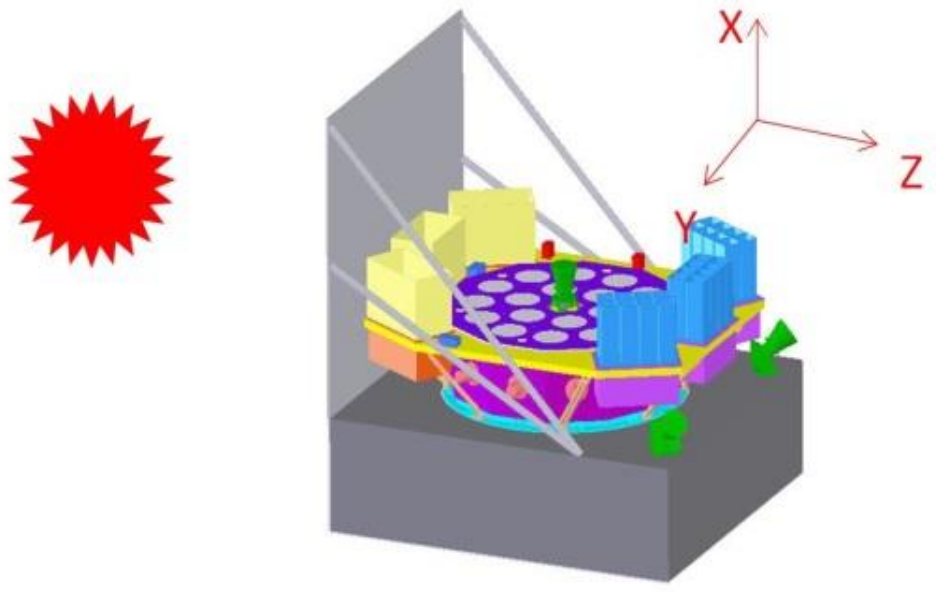

Figure 16 Thermal control of *Insight*-HXMT

### **4.6 ToO strategy and fast response of *Insight*-HXMT**

Once *Insight*-HXMT receives a ToO trigger from either the *Insight*-HXMT small area survey or from the astronomy society, the SGS will activate a procedure for fast response and arrange accordingly a ToO observation. The entire flow chart of *Insight*-HXMT ToO strategy is shown in Figure 17, from which one sees that *Insight*-HXMT can response to a ToO alert for fast response and to arrange a ToO observation within a time period as short as 5 hours.

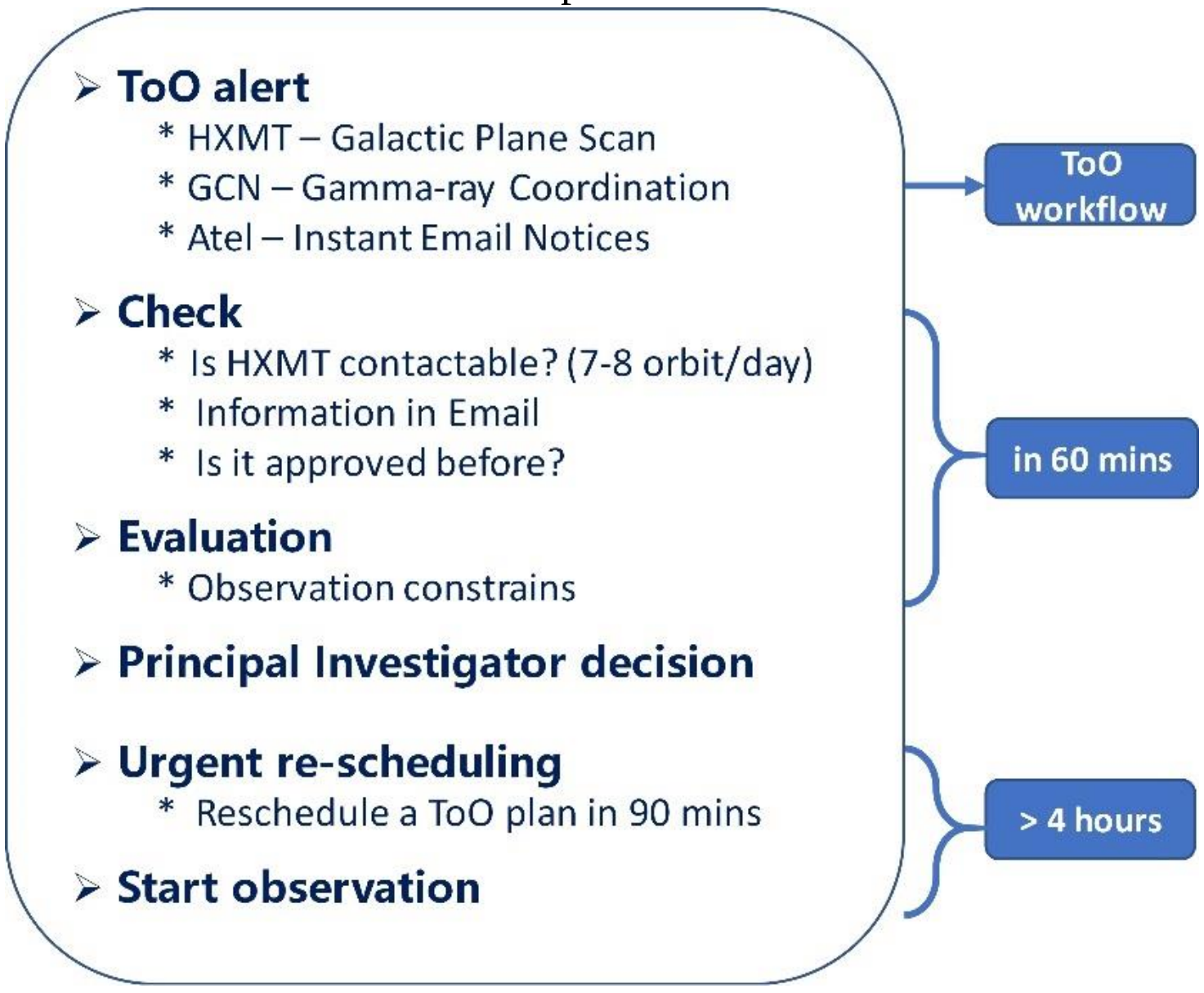

Figure 17 *Insight*-HXMT ToO response flow chart.

## 5. IN-ORBIT PERFORMANCE

*Insight*-HXMT was launched in JiuQuan satellite launch center at 11 am local time on June 15th 2017, and about 600 seconds later the launcher delivered successfully the satellite to the preset orbit (Figure 18). The platform was first tested in the next nine days, and so did the performances of the payload and the ground segment during the following three months. Shown in Figures 19-21 are a few examples for the in-orbit performances of HE, ME and LE: the dead time measurement of HE (Figure 19), the comparisons of the spectra measured in-orbit and on ground for pixels illuminated by the calibration radioactive sources (Figure 20 For ME and Figure 21 for LE). In Figure 22 we also show the global particle flux distribution measured by the particle monitors (PMs) during 2017 June 15 – July 11, which shows the SAA region clearly and thus verifies the functionality of PMs. The performance verification phase shows that the entire satellite work smoothly and healthily, and *Insight*-HXMT started to be in service for scientific observations at the end of 2017.

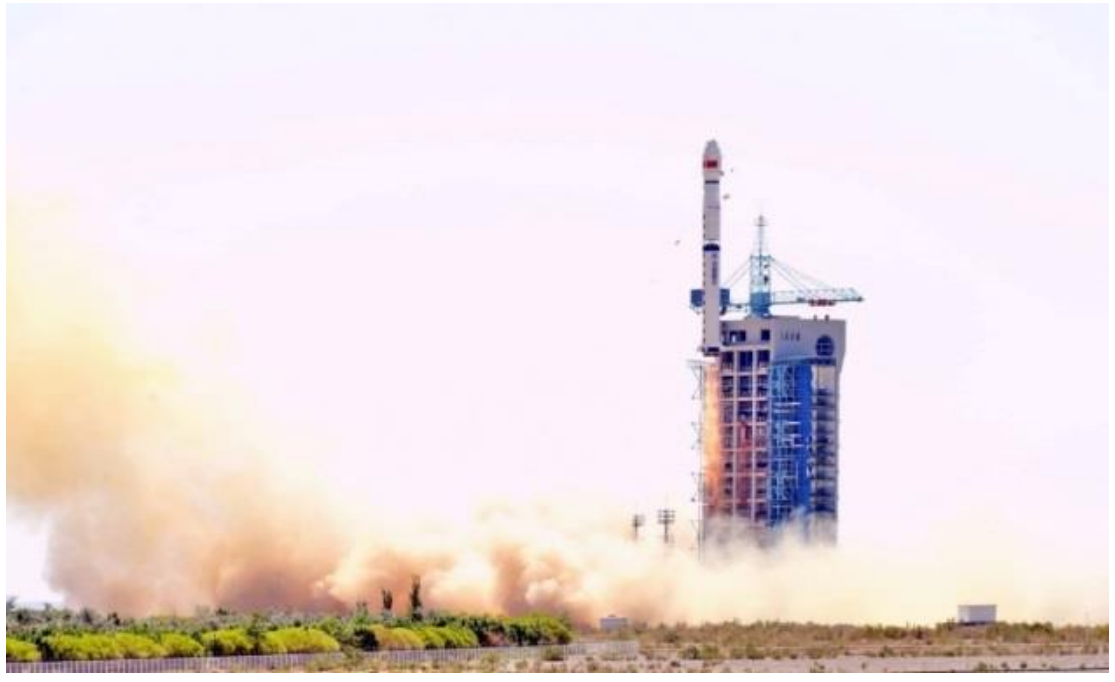

Figure 18 Launch of *Insight*-HXMT at Jiuquan Satellite Launch Center in northwestern China.

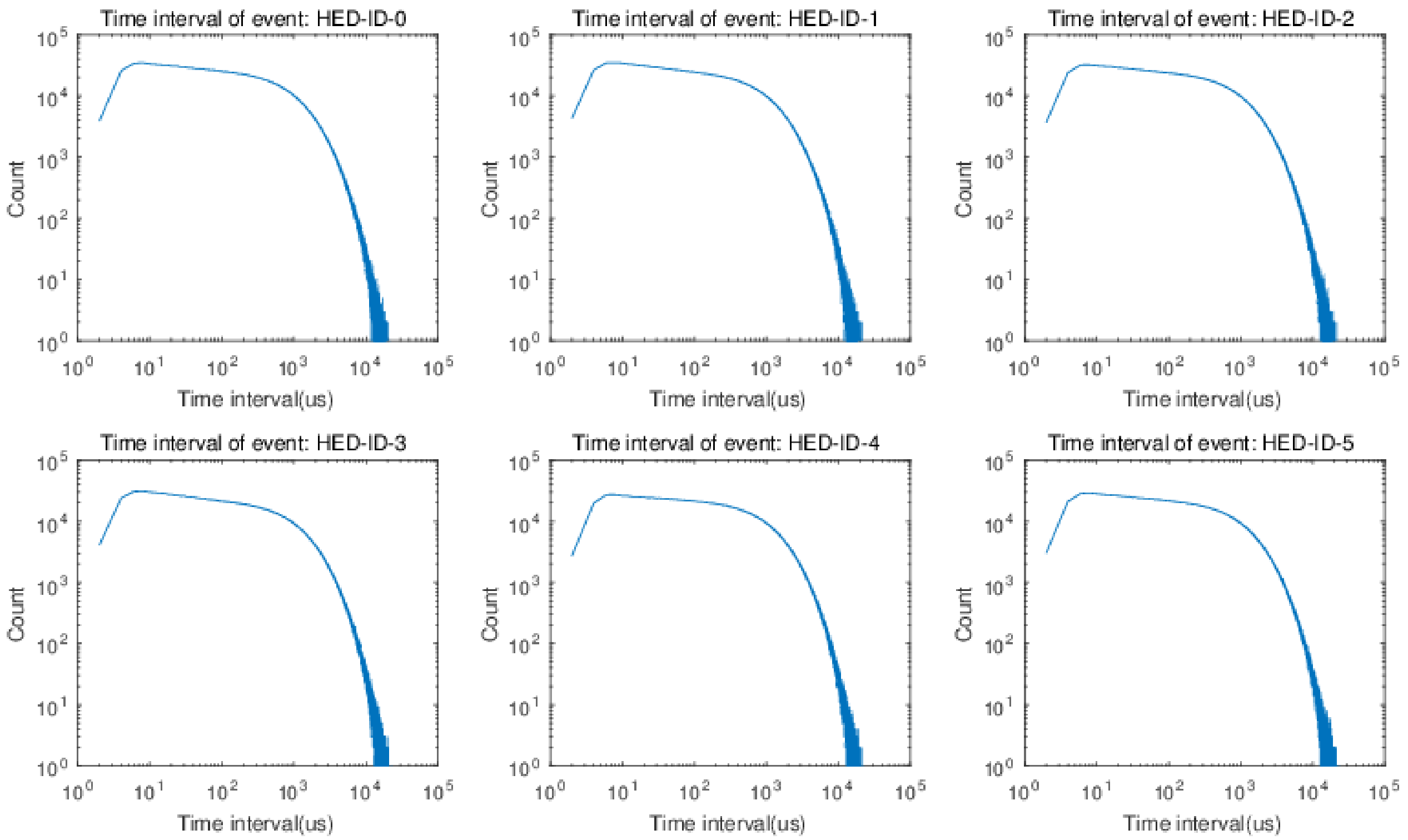

Figure 19 Measurements of the dead time for HE in orbit. In each panel the horizontal axis measures the arrival time interval of two sequentially recorded X-ray events and the vertical axis gives the corresponding count for each value of measured interval. Each panel corresponds to one readout channel for three HE detector units. The first break in the left part of each curve gives the dead time of the readout electronics of about 6 μs.

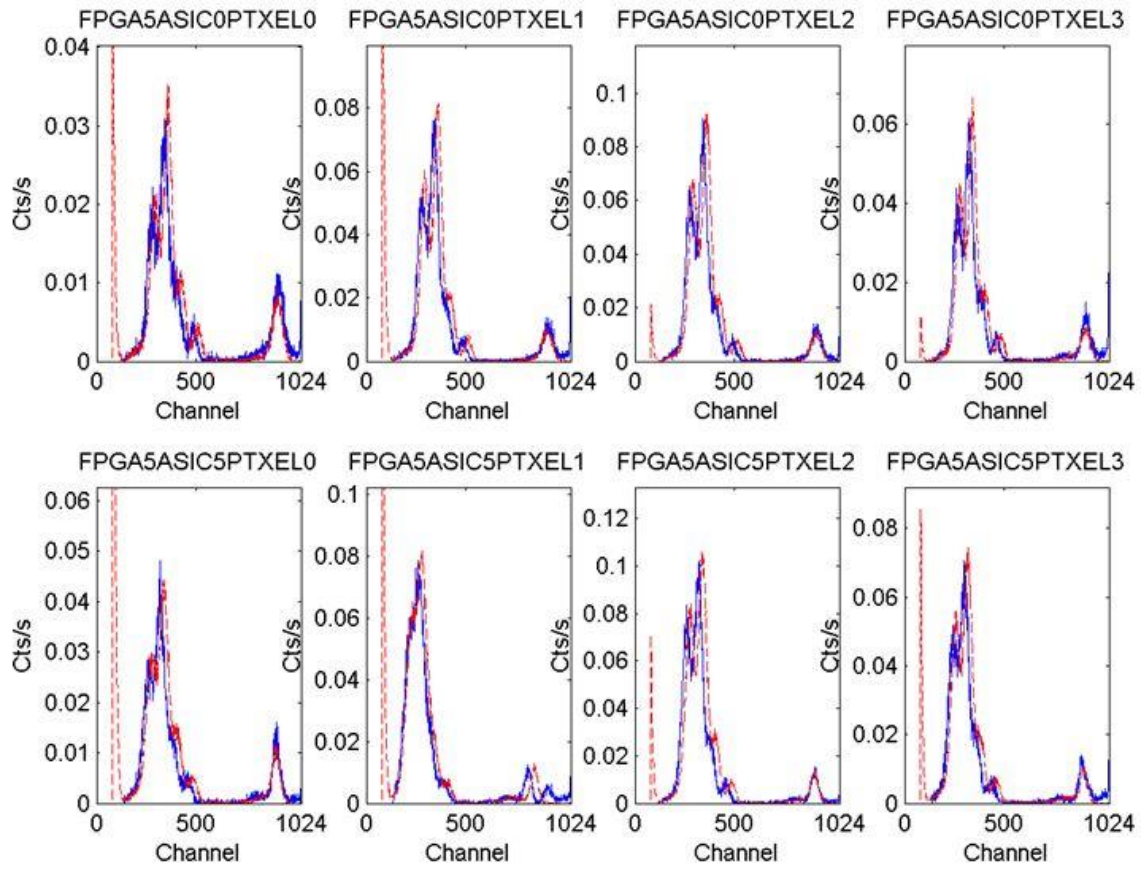

Figure 20 Comparisons of the spectra measured in-orbit and on ground for ME pixels illuminated by the calibration radioactive sources.

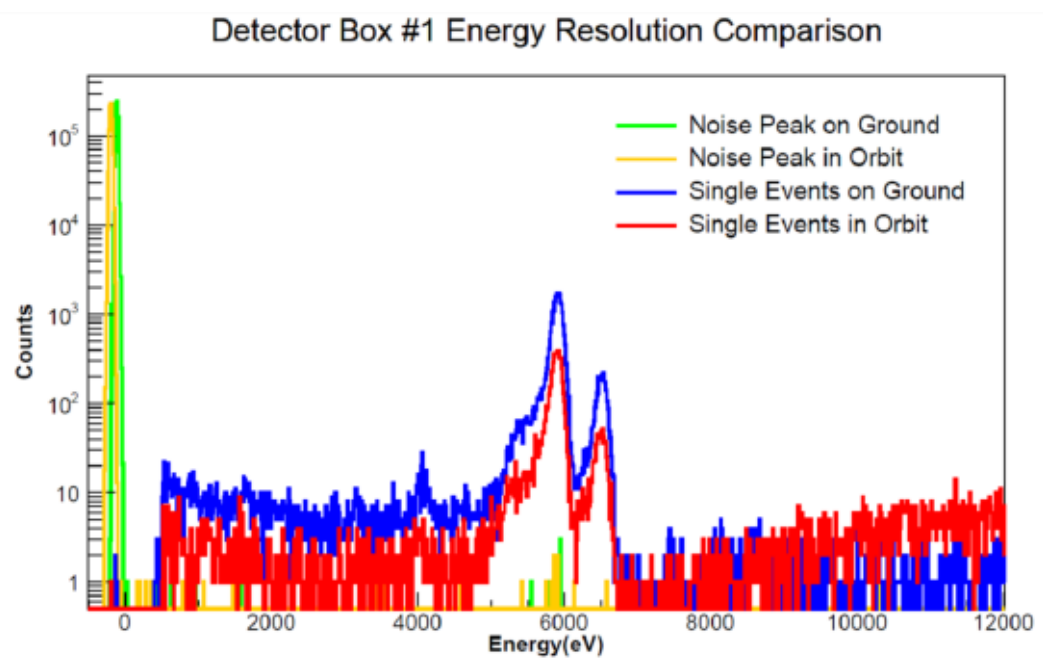

Figure 21 Comparisons of the spectra measured in-orbit and on ground for LE pixels illuminated by the calibration radioactive sources.

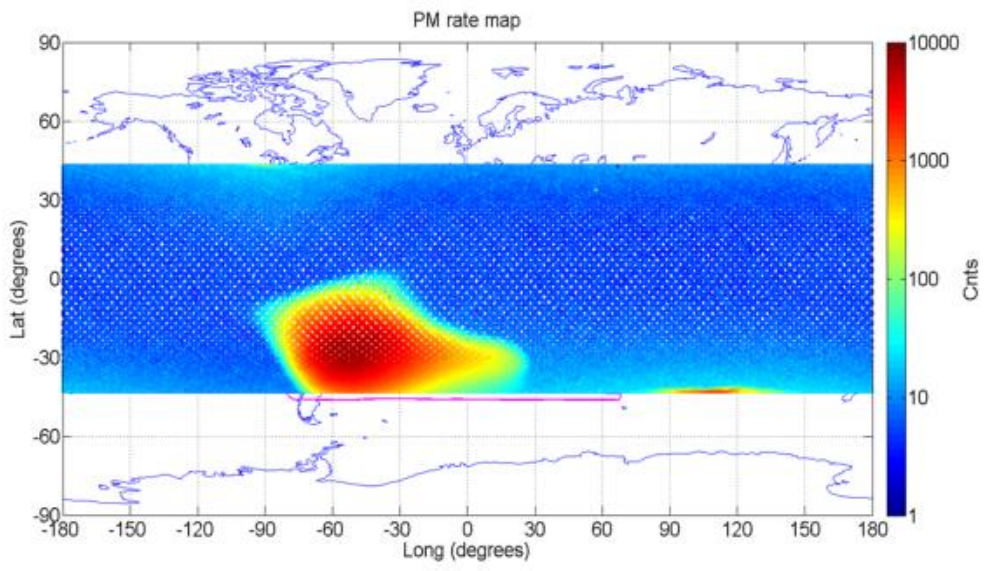

Figure 22 SAA region mapped by the *Insight*-HXMT particle monitors.

# 6 IN-ORBIT CALIBRATION

The *Insight*-HXMT in-orbit calibrations [25] include mainly the timing system, point spread function (PSF), energy response, effective area, and calibration in the GRB mode.

Calibration of the timing system needs the pulse-profile observed for isolated pulsars. Although the stable pulsar PSR 1509-58 was observed two weeks after the launch, it is too weak compared to the Crab pulsar to have sufficient statistics for timing calibration. The major calibrations were carried out when the Crab pulsar became visible after August 27 2017. As shown in Figure 23, the pulse profiles are derived for HE, ME and LE, which result in the time accuracy of 22 μs, 51 μs, and 21 μs for HE, ME and LE, respectively. The systematic shift in pulse profile of LE is due to the special read-out system of LE (Figure 24), which is well understood and hence the phase shift can be properly corrected.

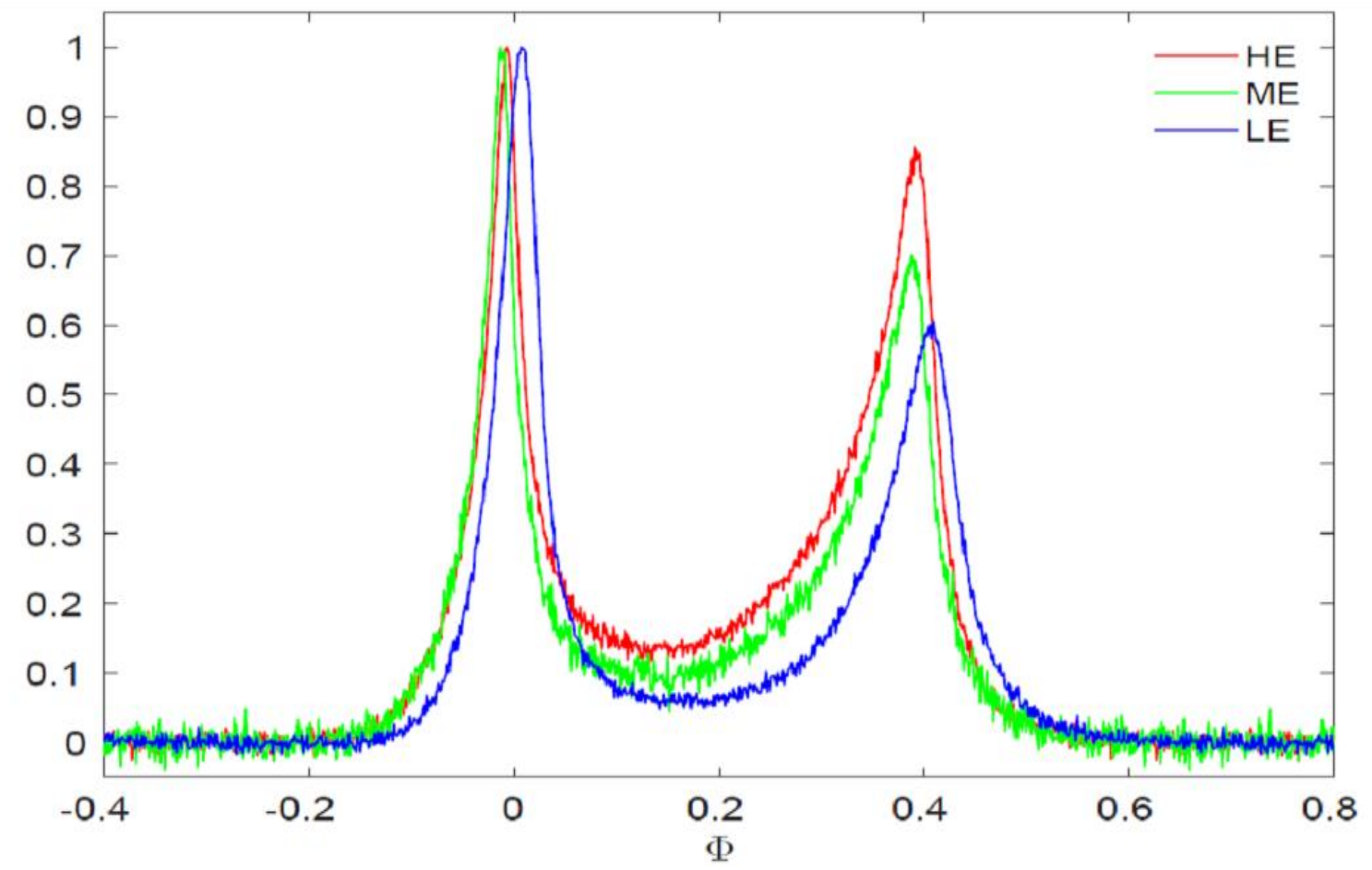

Figure 23 The pulse profile of the Crab pulsar as observed by HE, ME and LE.

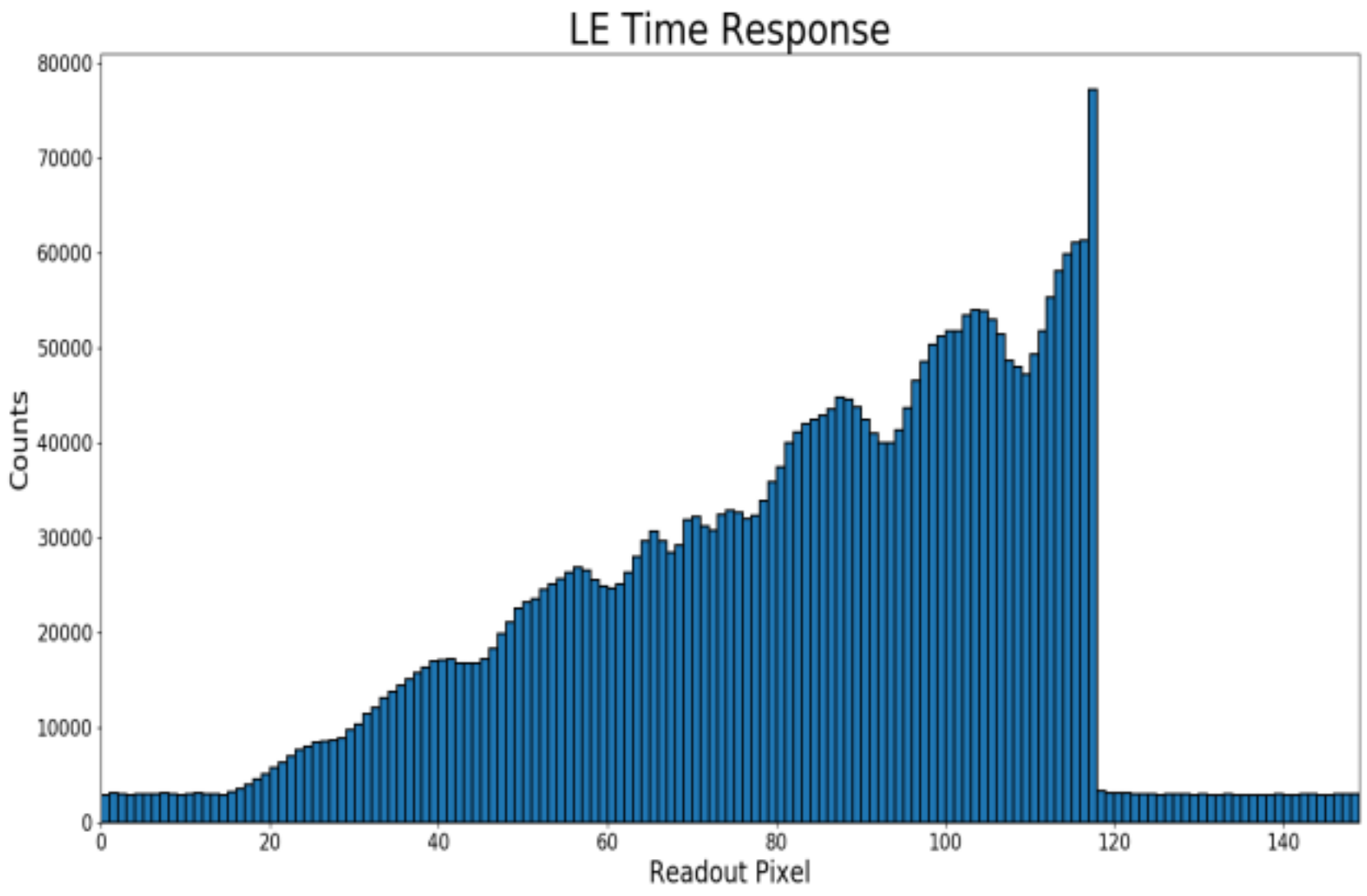

Figure 24 LE time response, which means that the recorded time of a photon is actually later than its arrival time up to 1 ms but with the exact time delay unknown. “Readout pixel” number is proportional to the geometrical distance between the charge readout location of each chip and the specific pixel hit by an incident X-ray photon.

The Crab pulsar and its nebula were observed in the scan mode for PSF calibration of *Insight*-HXMT. In the scan mode, Crab shows up in the light curve as a sequence of triangles (Figure 25). The PSF obtained on the ground was modified in order to fit well the source signal and reduce the residuals after PSF fitting (Figure 26).

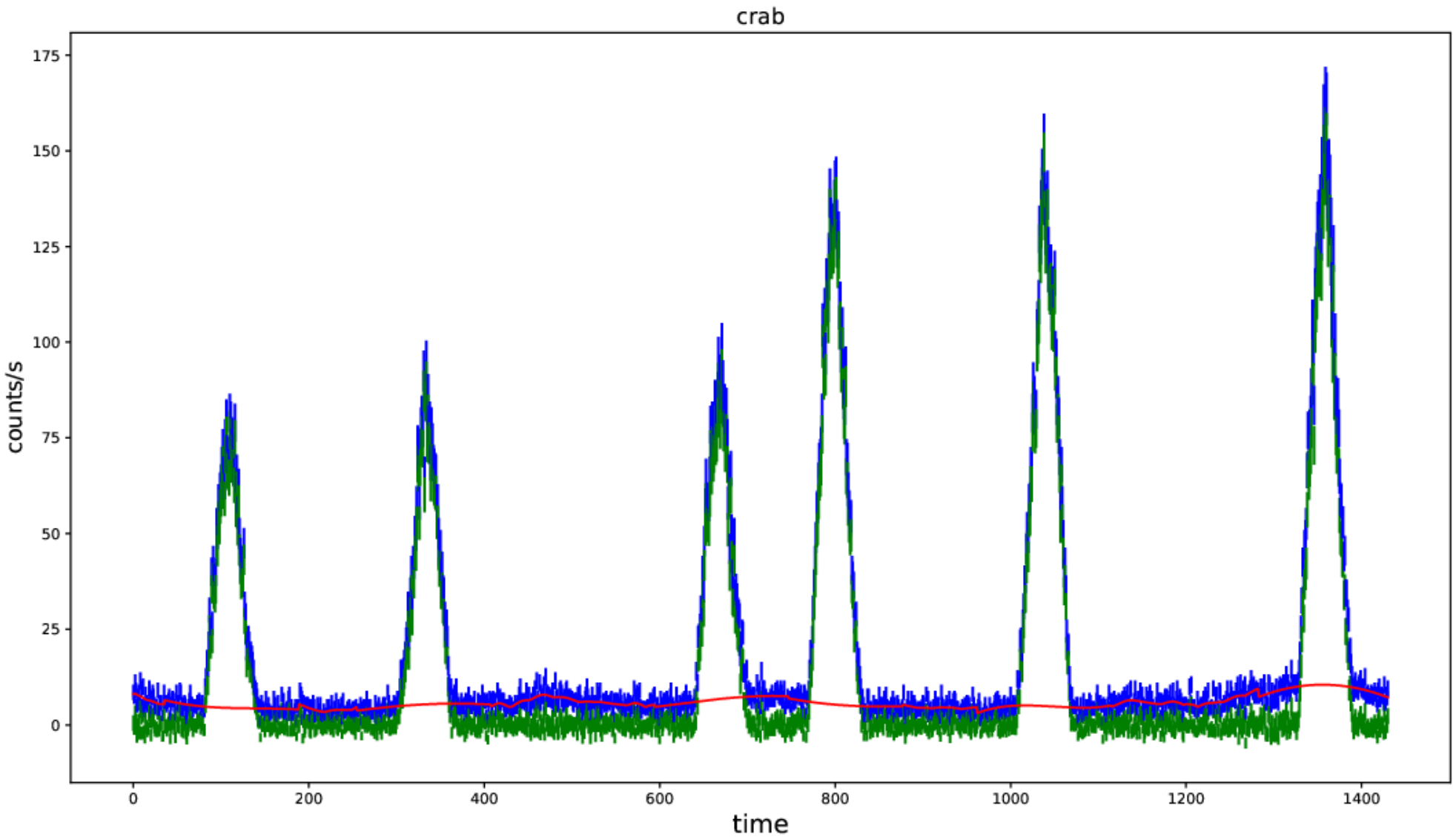

Figure 25 The light curve obtained by LE when it observed the Crab in a scanning observation. The red line shows the fit to the background that changes slowly. The observed data are shown in blue, and the green line shows the observed modulation of the Crab.

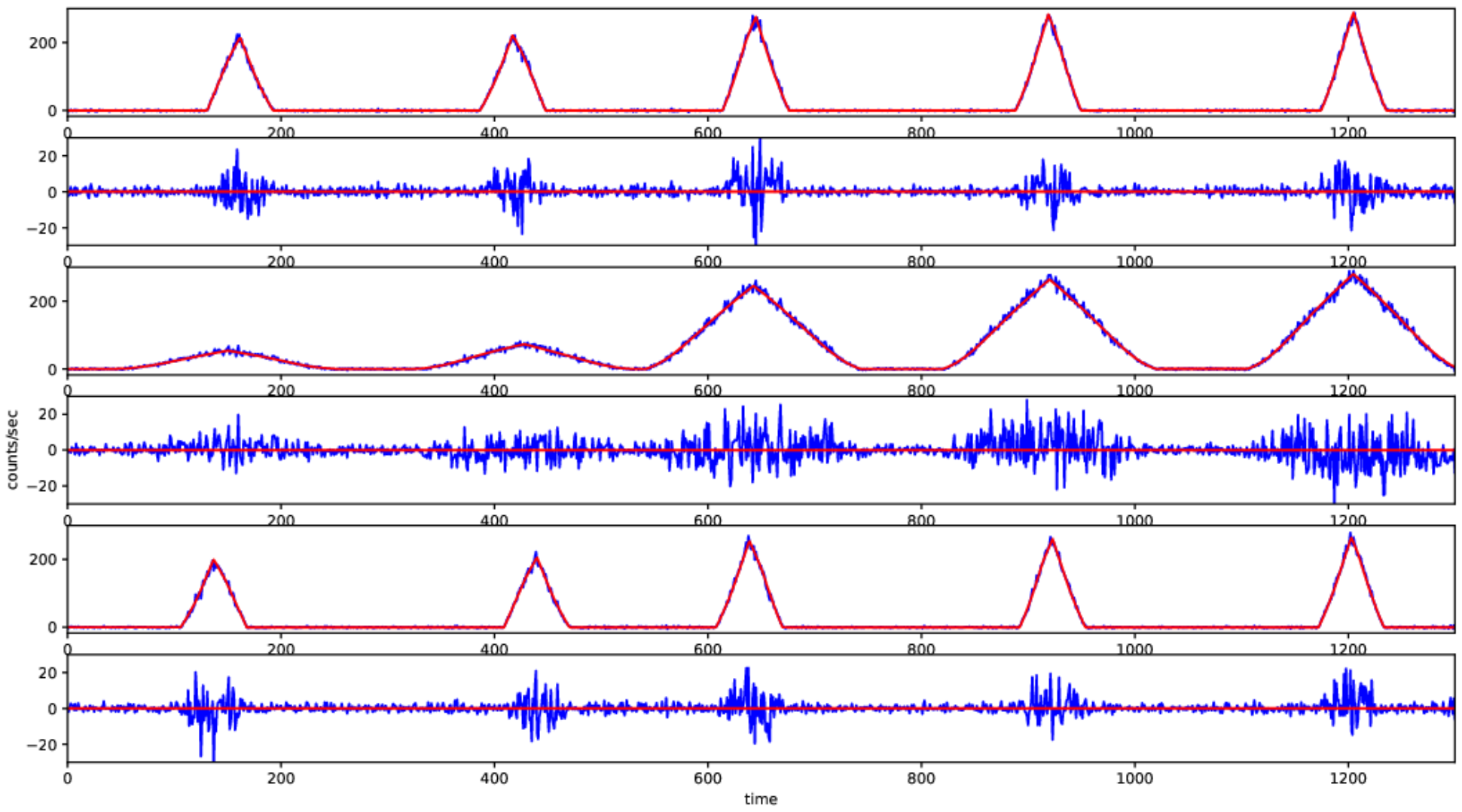

Figure 26 The light curves obtained by LE when it scanned across the Crab repeatedly. Each pair of plots denotes the light curve recorded by one LE detector box and the residual after PSF fitting to the source signal.

The calibration of the energy response consists of the energy-channel (EC) relation and the energy resolution. The energy resolution can be investigated by analyzing the line emission for pixels illuminated by the onboard radioactive sources and the results show that the energy resolution is consistent with that listed in Table 4. The calibration of the EC relation needs emission lines in the observed spectra and the calibrations are carried out in combination with the EC relation measured on the ground with the calibration facilities. For LE, the observations of supernova remnant Cas A provide six lines which are used to adjust the EC relation (Figure 27). For ME, only the silver line is visible in the background spectrum. Therefore, the overall property of the EC relation for ME is first evaluated from the more than 100 pixels that are illuminated by the onboard radioactive source $Am^{241}$ (Figure 28), and then compared to the EC relation measured on the ground. The EC relation for most of the ME pixels are adjusted based on their detection of the silver line. For HE, four lines from the background spectrum are used for the calibration of EC relation (Figure 29).

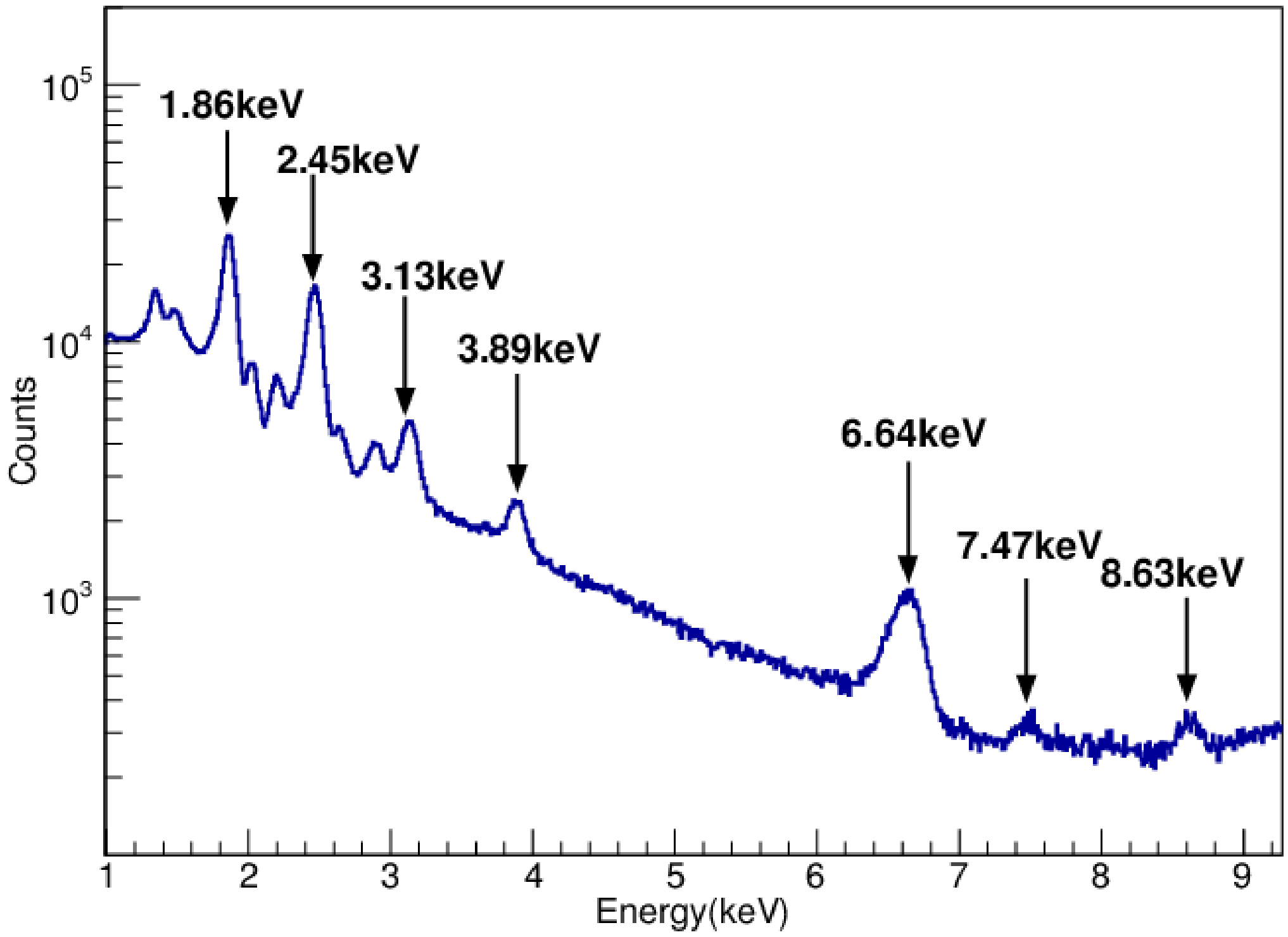

Figure 27 The X-ray spectrum of supernova remnant Cas A observed by LE

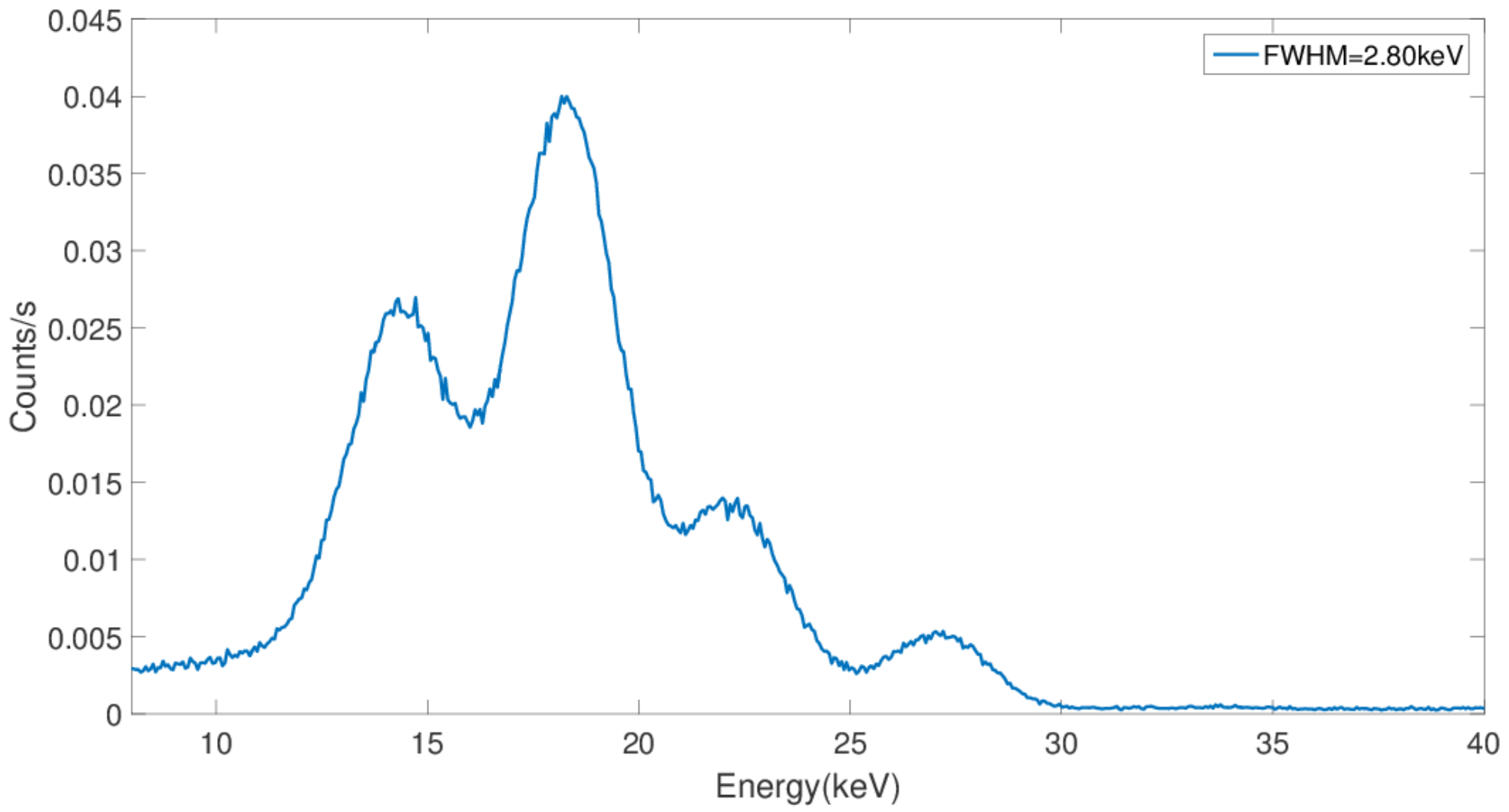

Figure 28 The spectrum of the onboard radioactive source $^{241}$Am obtained by the ME pixels illuminated.

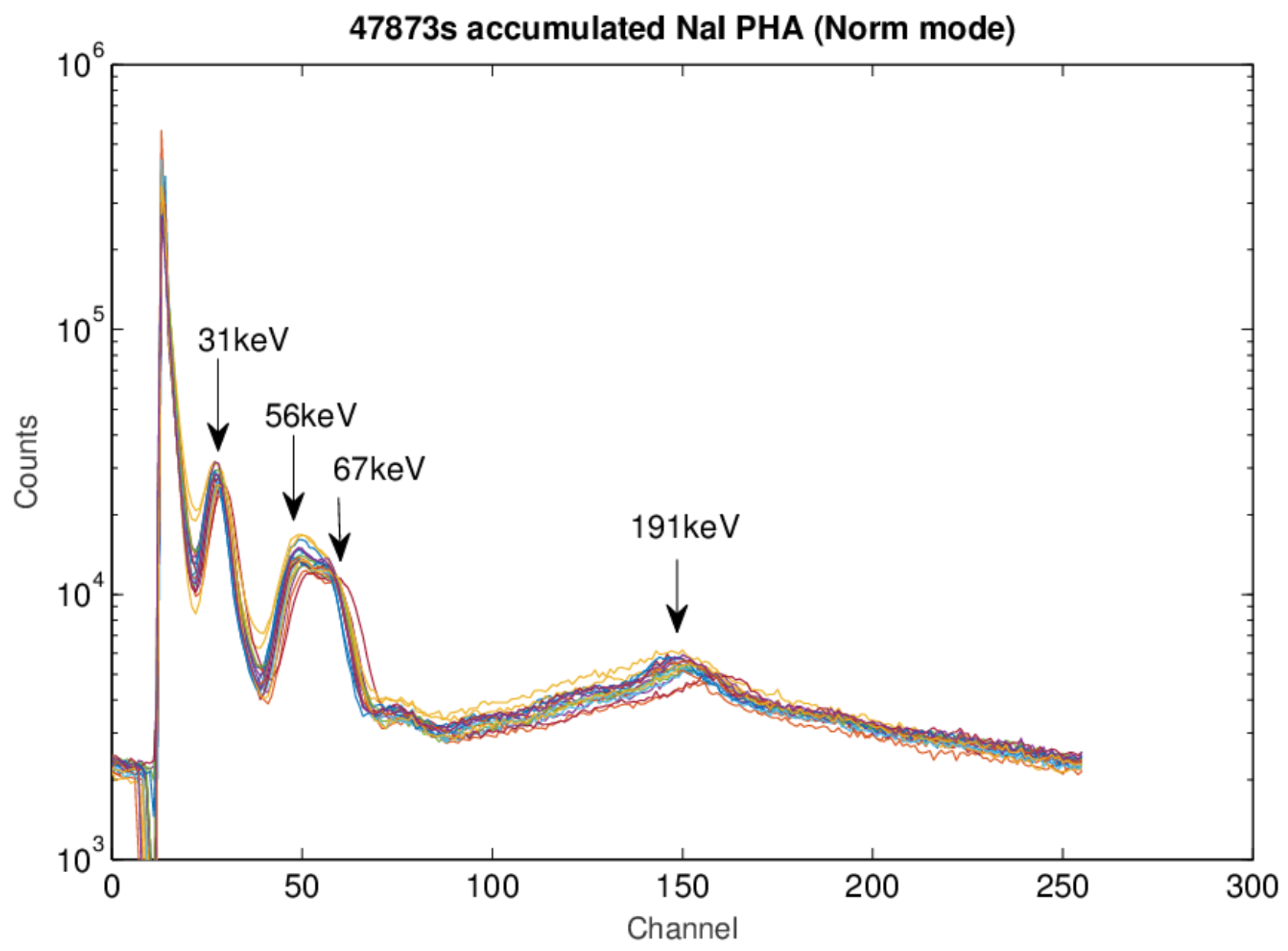

Figure 29 The background spectra of HE main detectors, which show four emission lines that are used to verify/adjust the EC relation of HE

The strategy of the calibration of the Ancillary Response File (ARF) is to take the known spectral shape of a source and to disentangle the uncertainty contributed by background modelling. For a collimated telescope the background modelling is usually a challenging and heavy task, which is currently ongoing in parallel to the ARF calibration for *Insight*-HXMT. We therefore take the pulse-off emission of the Crab pulsar as the background and calibrate the ARF with the pulse-on emission of this pulsar, which can eliminate the influence of the particle induced background that varies with time. The properties of the pulse-on spectrum are obtained mainly from the Rossi X-ray Timing Explorer (RXTE) observations of the Crab pulsar and is cross-checked for consistency with those obtained by BeppoSAX and NuSTAR. In the spectral fitting procedure, the spectral parameters of Crab's pulse-on emission are fixed, and the ARF calibration is performed via introducing an additional function to represent the residual of the spectral fitting. Fitting to the pulse-on spectrum of the Crab pulsar by using the final ARF files of *Insight*-HXMT is shown in Figure 30.

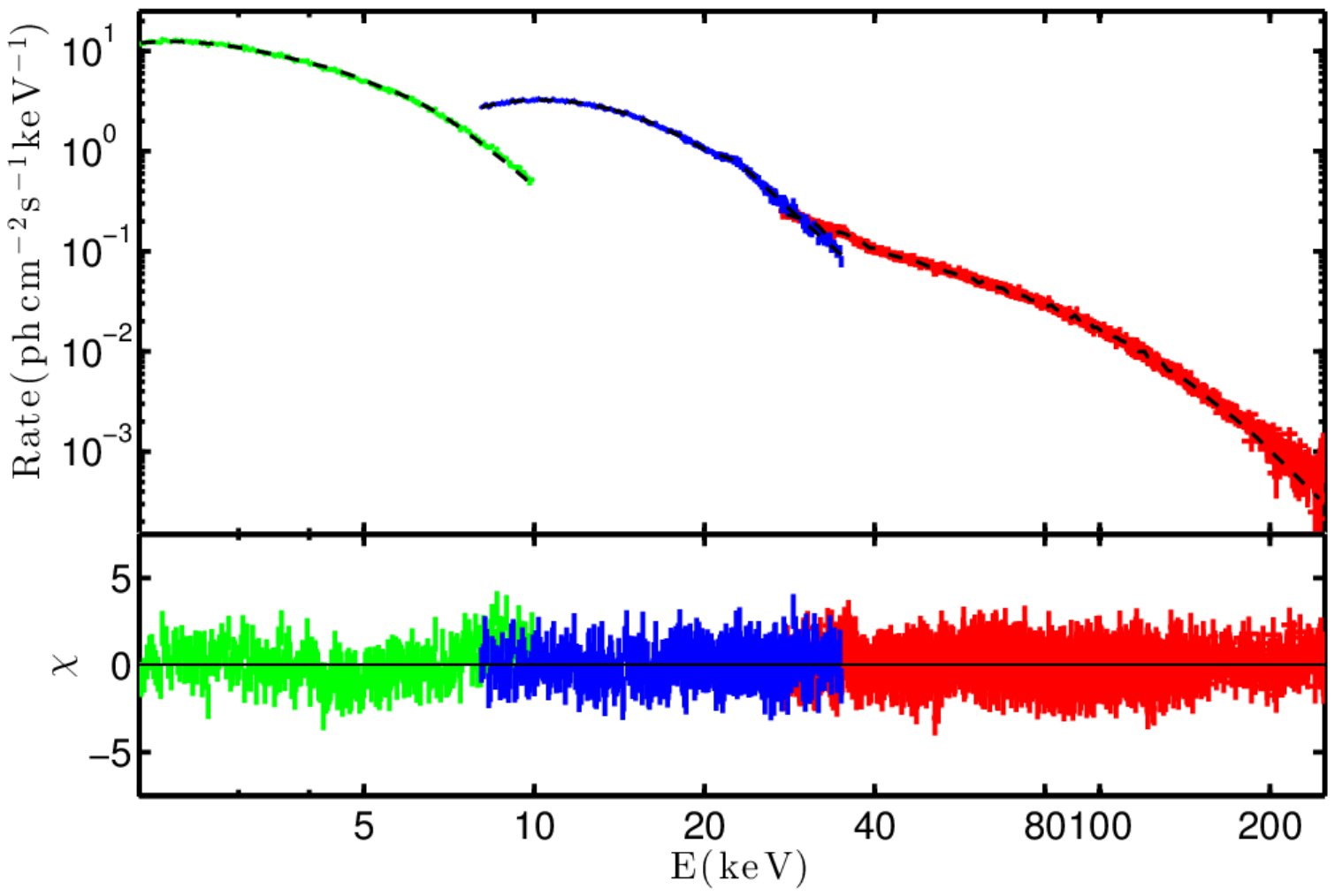

Figure 30 Fitting to the pulse-on spectrum of the Crab pulsar by using the final ARF files of HE, ME and LE.

The in-orbit calibration of CsI in the GRB mode turns out to be a complicated task. Some of the incident GRB photons are blocked by the platform structures and the CsI detectors can map photon distribution at a given incident direction of GRB event. The calibrations are carried out mainly based on the GEANT4 simulation and the observation of the Crab pulsar. The simulations show that, by using the structure of the platform as a coded mask, a typical GRB can be localized to an accuracy of a few degrees (Figure 31).

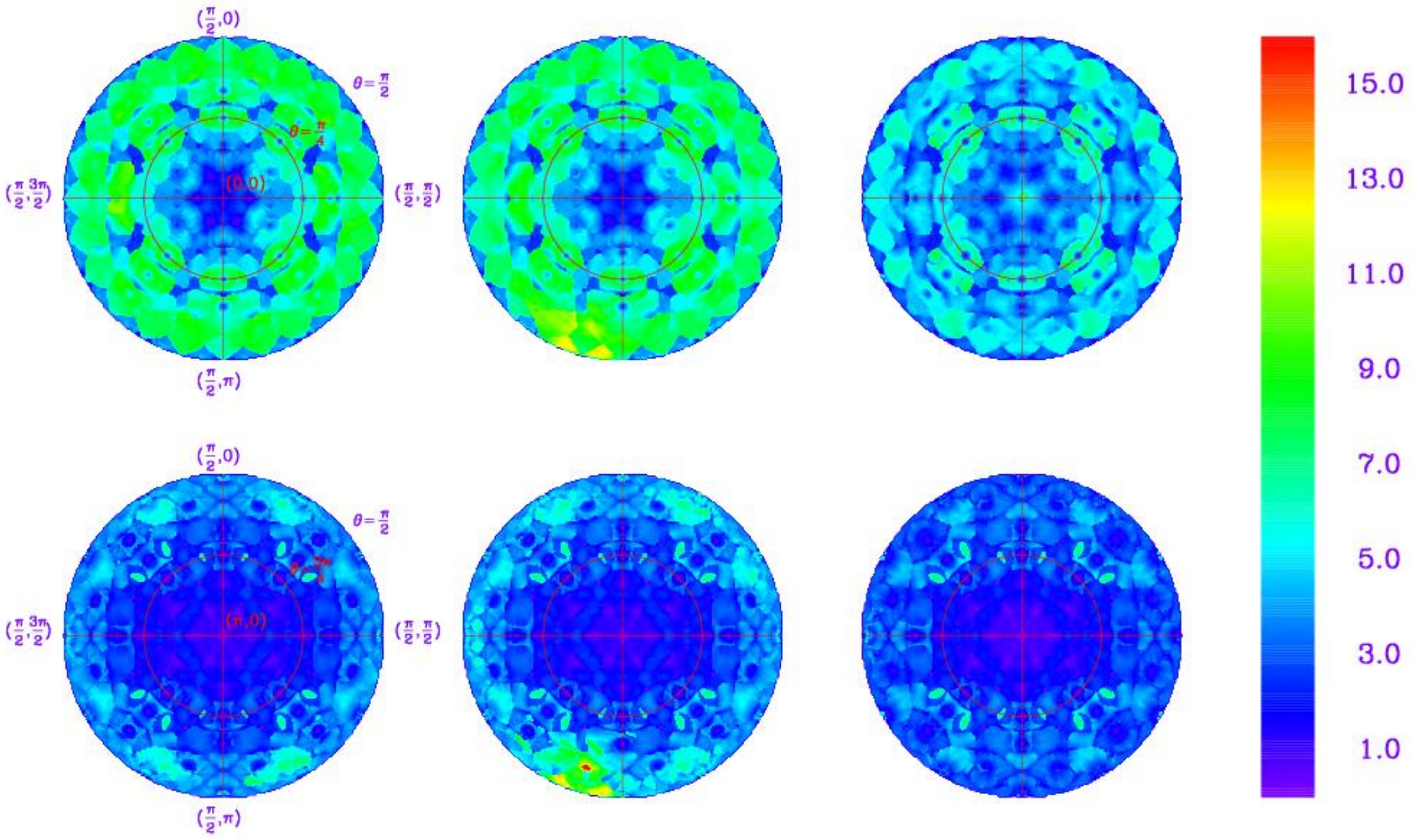

Figure 31 Localization capability of HE/CsI for GRBs. observation. Each map shows the localization error (color coded) as a function of the azimuthal and polar angles of the direction of a simulated GRB in the coordinates of the *Insight*-HXMT payload, for top incident (top three panels) and bottom incident (bottom three panels) GRBs. The three columns of maps correspond to three different GRB spectra described by the Band function with parameters (from left to right): α = (-1.9, -1.0, 0.0), β = (-3.9, -2.3, -1.5), $E_p$ = (70, 230, 1000) keV, fluence = (4x10$^{-5}$, 2x10$^{-5}$, 1x10$^{-5}$) erg/cm$^2$, respectively; the integration time is 10 s for all cases.

## 7. IN-ORBIT BACKGROUND

For a collimated telescope the background is usually measured using the conventional on/off observational mode. This is, however, not possible for *Insight*-HXMT, since it was initially designed only for scanning survey and not for pointed observations. We therefore changed slightly the original mission design by blocking the collimators of some detectors completely, to estimate the instrumental background. Accordingly, the background model is developed by taking the correlations of the count rates between the blind and other detectors. Observations of a series of blank sky regions are used as input for building and testing the background model. Figure 32 demonstrates how the background of a blank sky observation is recovered from the background model built for one HE module.

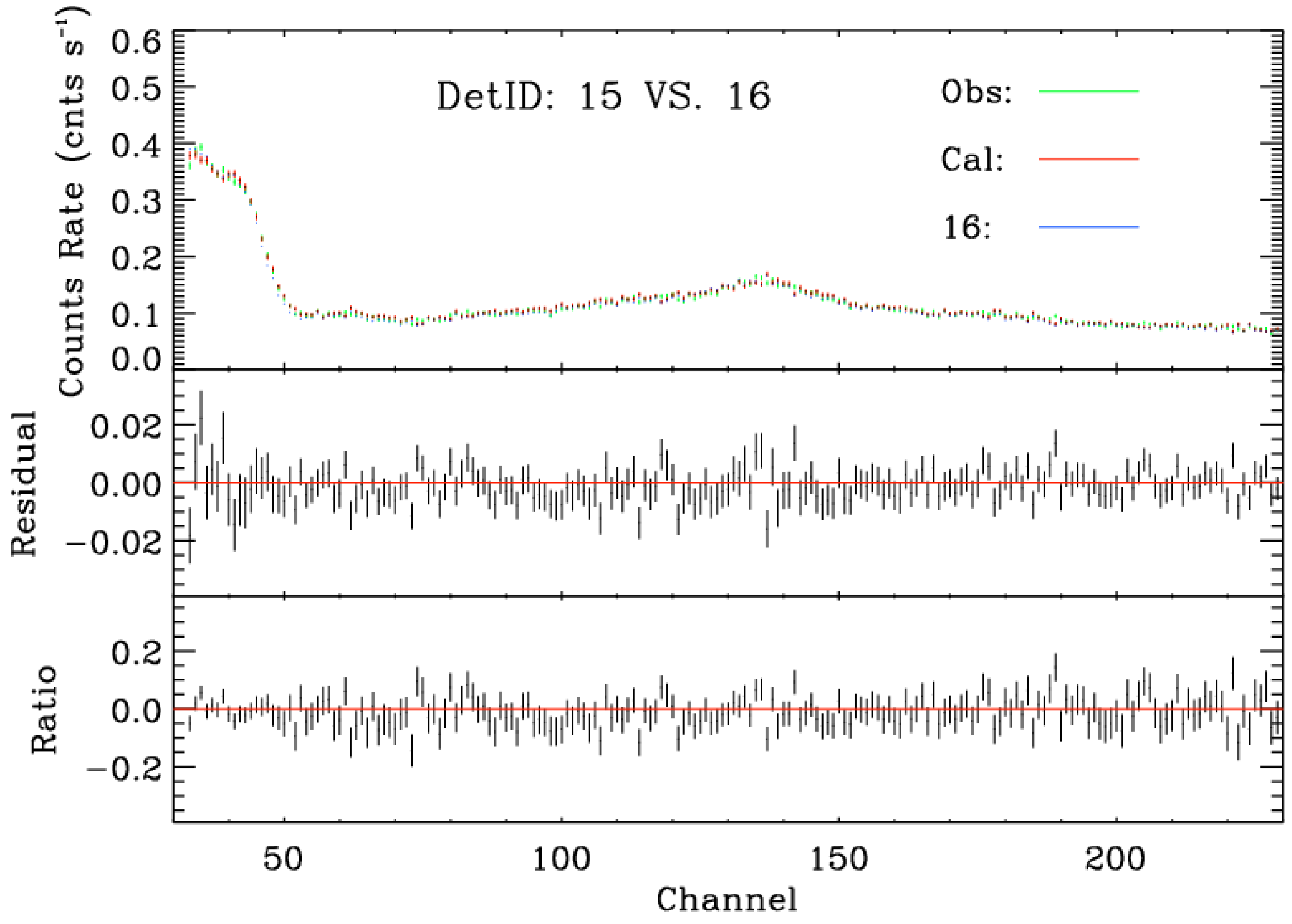

Figure 32 The HE spectrum of a blank sky region observed by HE module No.15. In the upper panel, the green points are the observational data, the blue points are the contemporary background measured by the blind module No.16, and the red points represent the background estimated based on the background model. The two lower panels show residual of the spectral fit and the ratio between the residuals and the real data of HE module No.15.

## 8. OBSERVATIONS

Till 2018 June, *Insight*-HXMT has been in service for one year. A variety of observations have been carried out by *Insight*-HXMT, which include over than 680 pointings and roughly 460 scanning surveys. A summary is given in Table 5 in detail while the sky explored so far by *Insight*-HXMT is shown in Figure 33.

Table 5: Summary of the HXMT observations in the first year

| | **Mode** | **Type** | **Source Name** | **Pointings** | **Exposure (s)** |
|---|---|---|---|---|---|
| 1 | Point | Supernova remnant | Cas A | 9 | 530 |
| 2 | | Pulsar | Crab | 76 | 1230 |
| 3 | | | PSR B0540-69 | 7 | 250 |
| 4 | | | PSR B1509-58 | 12 | 310 |
| 5 | | BH Binary | Cyg X-1 | 13 | 300 |
| 6 | | | Granat 1716-249 | 2 | 250 |
| 7 | | | GRS 1915+105 | 24 | 720 |
| 8 | | | GX 339-4 | 1 | 100 |
| 9 | | | H 1743-322 | 15 | 180 |
| 10 | | | MAXI J1535-571 | 18 | 430 |
| 11 | | | MAXI J1543-564 | 1 | 80 |
| 12 | | | MAXI J1727-203 | 3 | 30 |
| 13 | | | MAXI J1820+070 | 61 | 1360 |
| 14 | | | Swift J1658.2-4242 | 23 | 470 |
| 15 | | | Cyg X-3 | 16 | 410 |
| 16 | | NS Binary | 2A 1822-371 | 1 | 30 |
| 17 | | | 4U 1728-34 | 4 | 90 |
| 18 | | | 4U 0115+63 | 11 | 150 |
| 19 | | | 4U1636-536 | 24 | 250 |
| 20 | | | Aql X-1 | 3 | 30 |
| 21 | | | Cen X-3 | 14 | 400 |
| 22 | | | Cir X-1 | 6 | 100 |
| 23 | | | Cyg X-2 | 24 | 570 |
| 24 | | | GRO J1008-57 | 11 | 340 |
| 25 | | | GRO1750-27 | 1 | 15 |
| 26 | | | GS 1826-238 | 1 | 40 |
| 27 | | | GX 301-2 | 15 | 400 |
| 28 | | | GX9+9 | 5 | 110 |
| 29 | | | GX 13+1 | 1 | 30 |
| 30 | | | GX 17+2 | 10 | 230 |
| 31 | | | H 1417-624 | 21 | 210 |
| 32 | | | Her X-1 | 16 | 470 |
| 33 | | | IGR J16328-4726 | 2 | 20 |
| 34 | | | NGC 6624 | 1 | 30 |
| 35 | | | PSR J2032+4127 | 4 | 40 |
| 36 | | | Sco X-1 | 7 | 230 |
| 37 | | | Swift J1756.9-2508 | 1 | 40 |
| 38 | | | Swift J0243.6+6124 | 97 | 1200 |
| 39 | | | Vela X-1 | 1 | 120 |
| 40 | | | XTE J1946+274 | 1 | 100 |
| 41 | | Extra-galactic | 1ES 1959+650 | 25 | 255 |
| 42 | | | Cosmos Field | 4 | 80 |
| 43 | | | M87 | 4 | 180 |
| 44 | | | Perseus | 2 | 200 |
| 45 | | Blank Sky | 21 regions | 87 | 870 |
| 46 | Small Area Scan (SAS) | Crab Area | around the Crab | 9 | 550 |
| 47 | | Galactic Plane | 22 regions | 458 | 5000 |

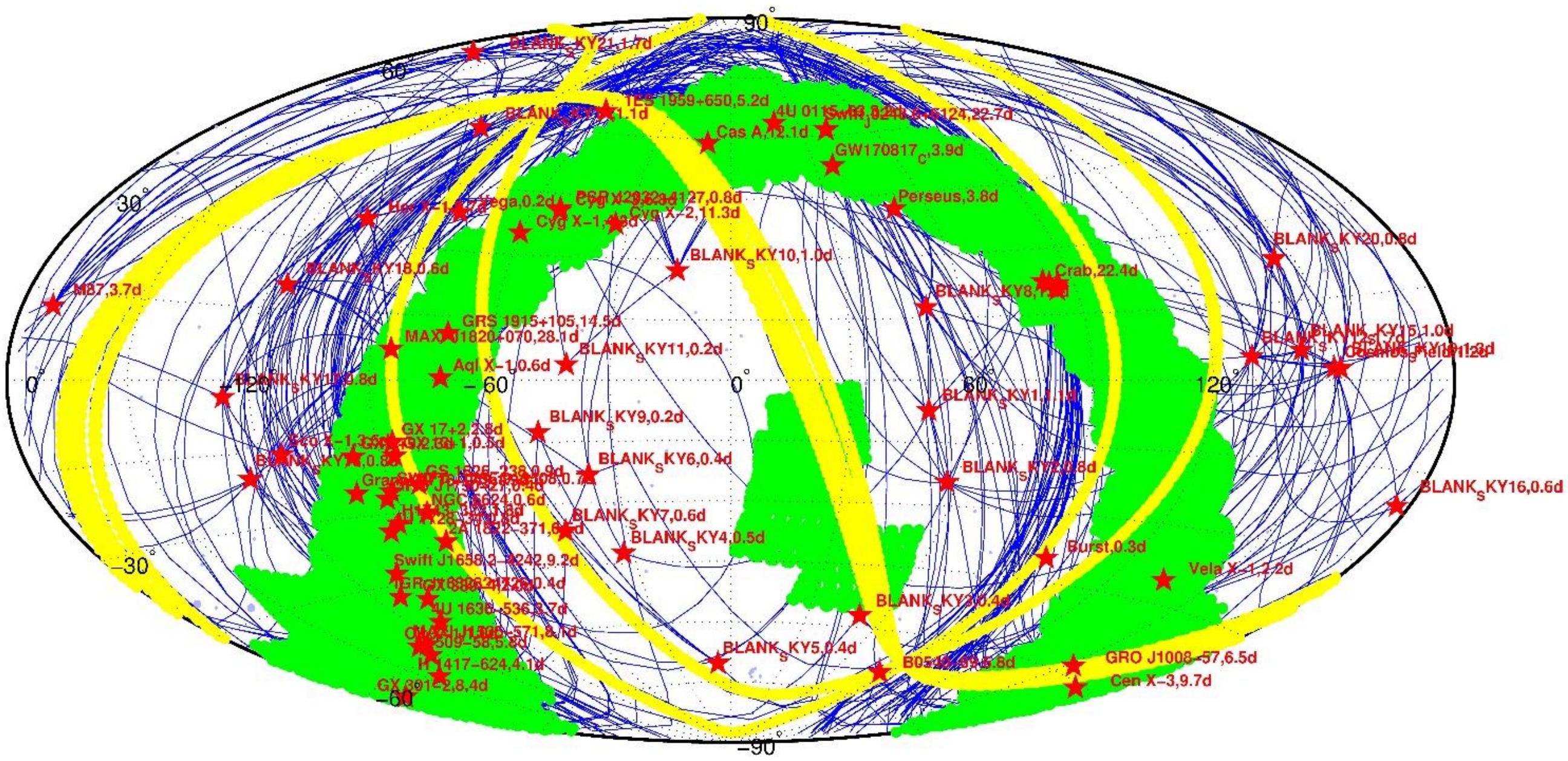

Figure 33 Sky coverage of *Insight*-HXMT till 2018 April. Here the red stars mark the pointed observations, the green belts are the small area scans, the yellow belts are the test of the all-sky survey mode, blue lines are tracks of slew between different observations.

## 9. PRELIMINARY RESULTS

The preliminary results obtained by *Insight*-HXMT cover a rather wide scope: Galactic plane survey, timing and spectral analyses of outbursting BH XRB and NS XRB systems, timing and spectroscopy of pulsars, bursts from low mass XRBs, observation of EM counterpart of the NS-NS merger event GW170817, GRBs, etc. Here we only show some examples.

Shown in Figure 34 is the Galactic center region mapped by applying the DDM imaging method to LE observation. Figure 35 shows the low frequency QPOs detected by HE, ME and LE in the newly discovered BH XRB transient MAXI J 1535-571 [26]. For the newly discovered NS XRB transient Swift J0243.6+6124, the pulse profile of the neutron star shows complicated patterns along with the outburst evolution (Figure 36). During the famous GW EM observational campaigns, *Insight*-HMXT was among the four X/gamma-ray telescopes that monitored the GW source throughout the trigger time (Figure 37) [27], and hence reported the most stringent constraints to the emission at soft gamma-rays (Figure 38) for GW170817 during time periods of the precursor, the GBM trigger, and the afterglow[28]. More than 60 GRBs were detected by HXMT/HE until April 2018 and Figure 39 shows one example, demonstrating the good sensitivity of HE for the detection of short/hard GRBs. Other scientific results of Insight-HXMT can be found from these references [29,30,31]

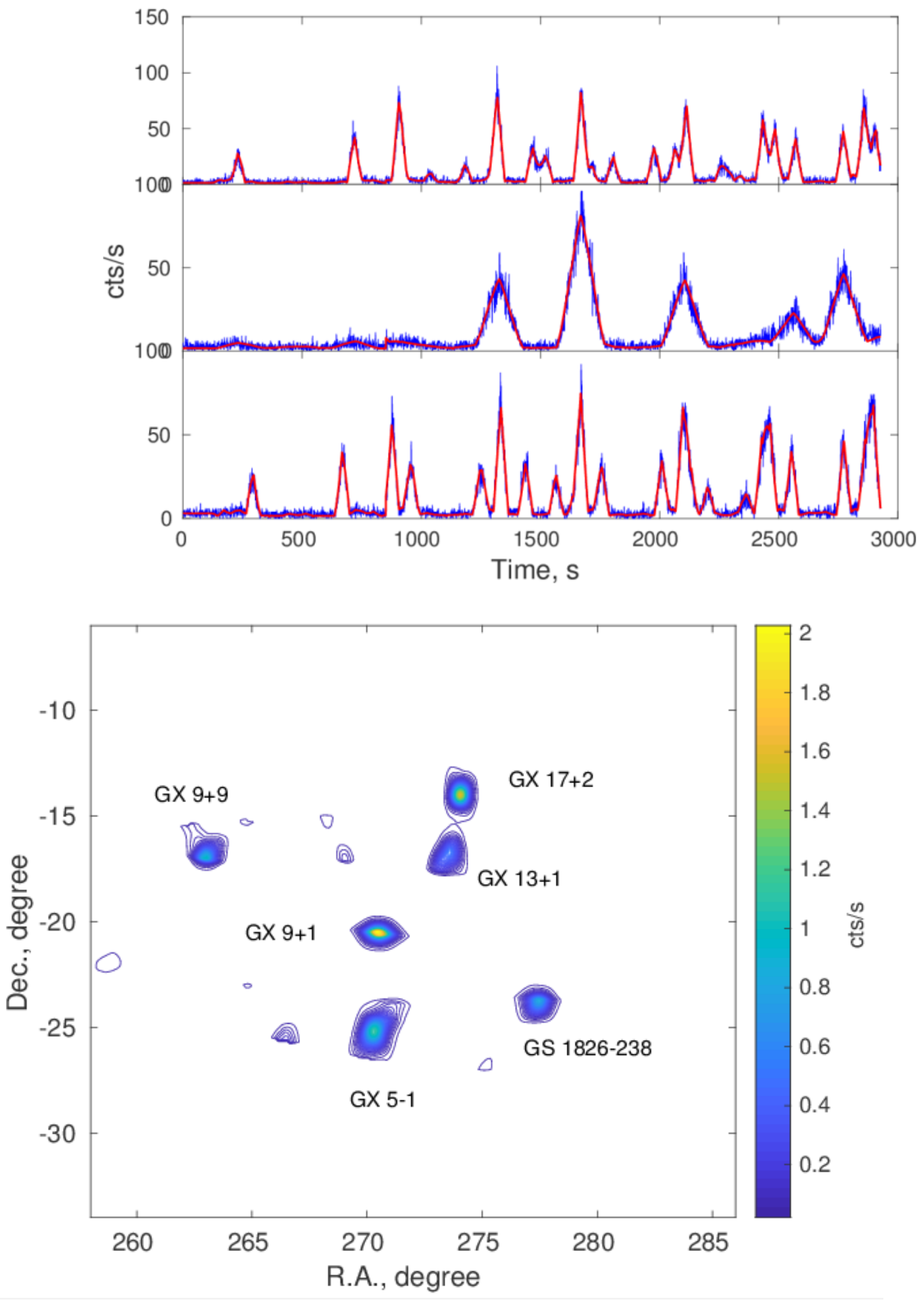

Figure 34 LE observation of the Galactic cernter region: the upper and lower panels are the scan light curves and sky map reconstracted from those light curves with the direct demodulation method.

Table 6: List of identified sources in the reconstructed map of the Galactic center region shown in Fig. 34.

| Source name | Reconstructed Position | | Real Position | | Error (degree) |
|---|---|---|---|---|---|
| | Ra (degree) | Dec (degree) | Ra (degree) | Dec (degree) | |
| GX 9+9 | 270.281 | -24.961 | 270.284 | -25.079 | 0.12 |
| GX 9+1 | 270.440 | -20.553 | 270.385 | -20.529 | 0.06 |
| GX 5-1 | 274.022 | -13.979 | 274.006 | -14.036 | 0.06 |
| GX 17+2 | 277.339 | -23.858 | 277.367 | -23.797 | 0.07 |
| GX 13+1 | 273.595 | -17.072 | 273.631 | -17.157 | 0.09 |
| GS 1826-238 | 262.973 | -16.858 | 262.934 | -16.962 | 0.11 |

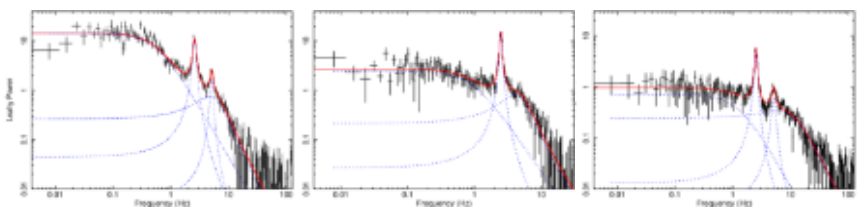

Figure 35 The low frequency QPOs detected by HE (right), ME (middle) and LE (left) under an exposure of 3ks, during the low/hard state of the outburst from the newly discovered BH candidate MAXI J1535-571 [13].

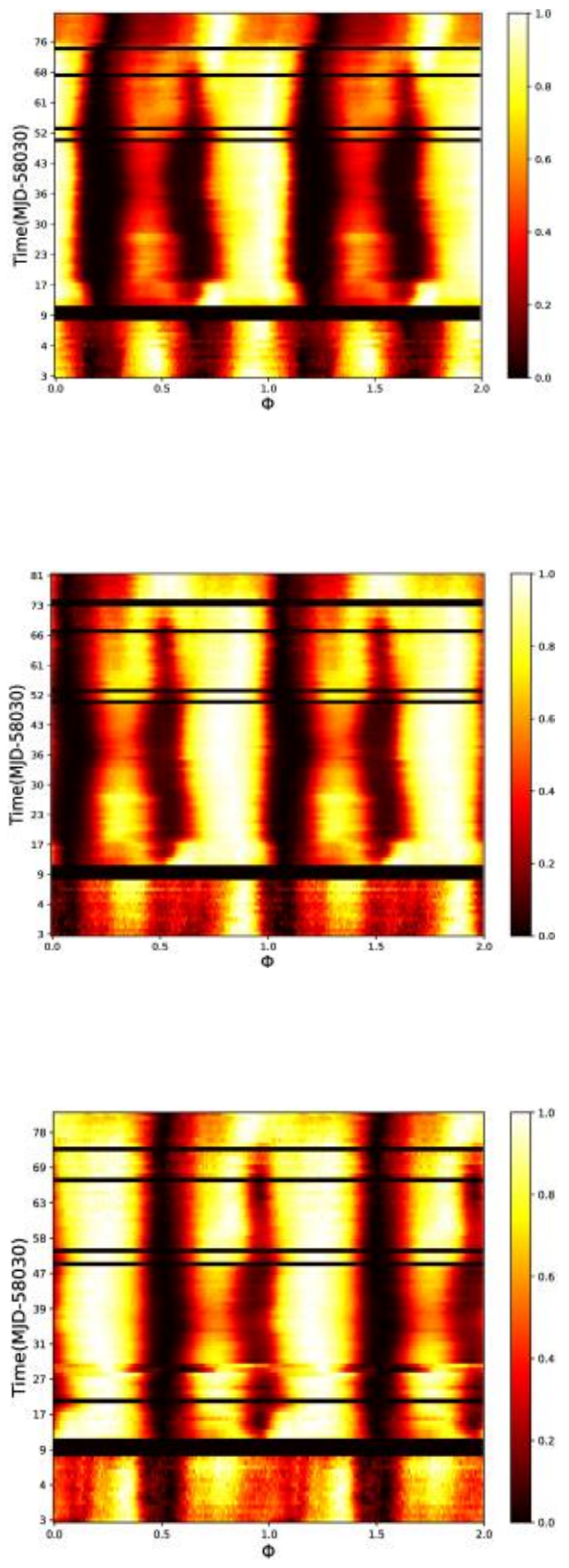

Figure 36 Pulse profile evolution of the NS star harbored in the newly discovered NS XRB system Swift J0243.6+614 during the outburst. From top to bottom are the pulse profiles recorded by HE, ME and LE, respectively.

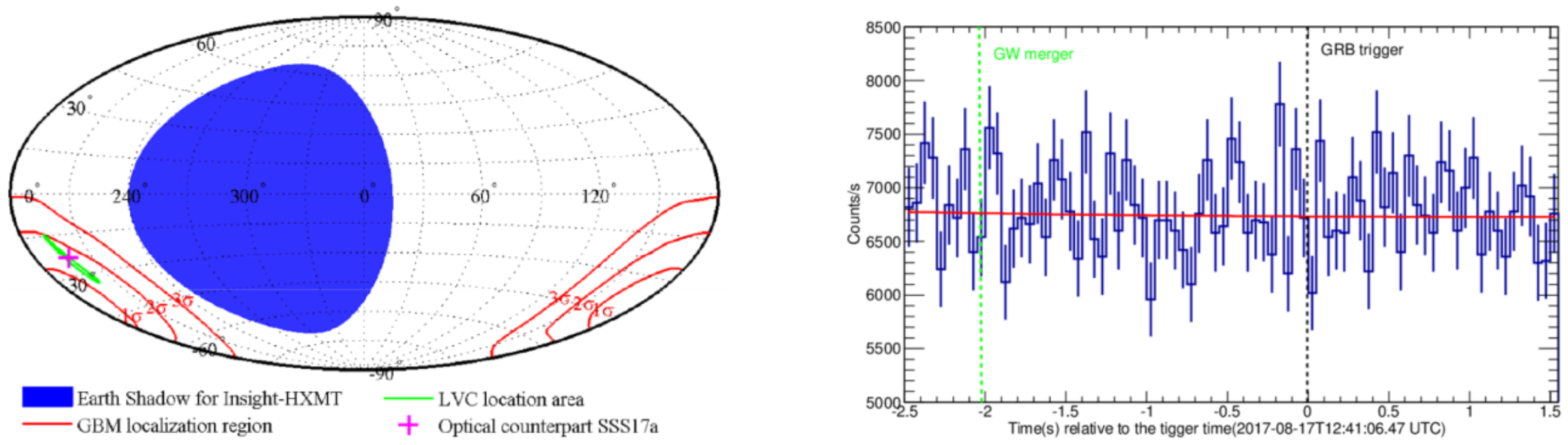

Figure 37 The sky region of GW170817 monitored by *Insight*-HXMT/CsI (left), and the light curve (right) covers both time perods around the GW merger and GRB trigger.

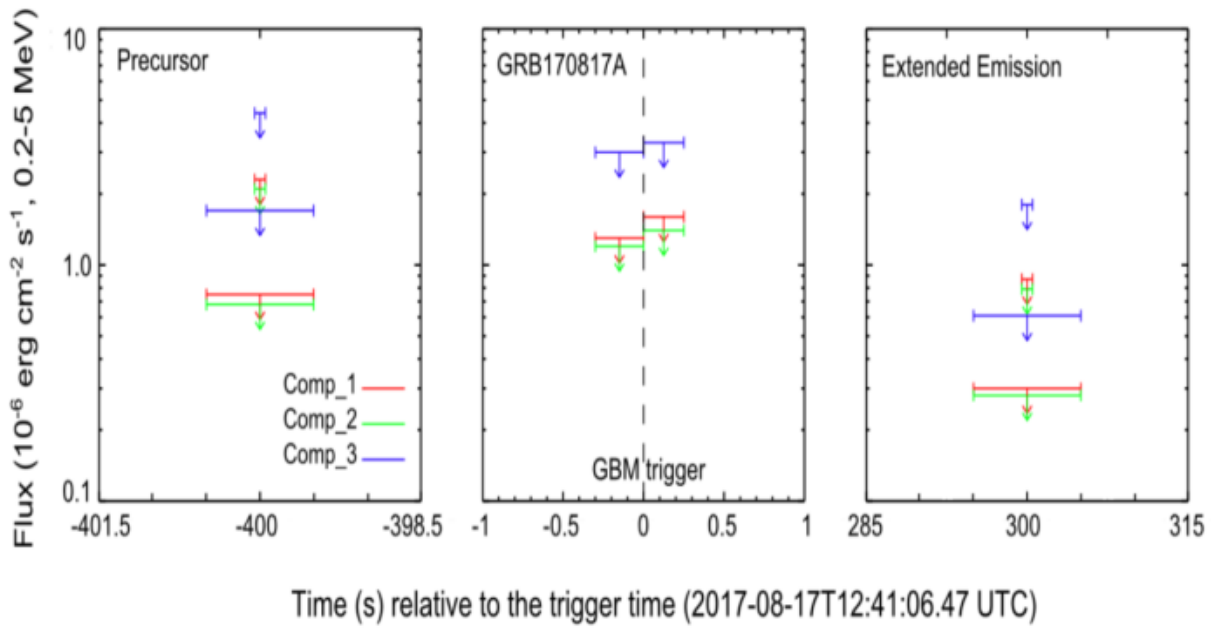

Figure 38 The *Insight*-HXMT detections of the flux limits of the EM counterpart of GW170817 during precursor, GRB trigger and for the extended emission [9].

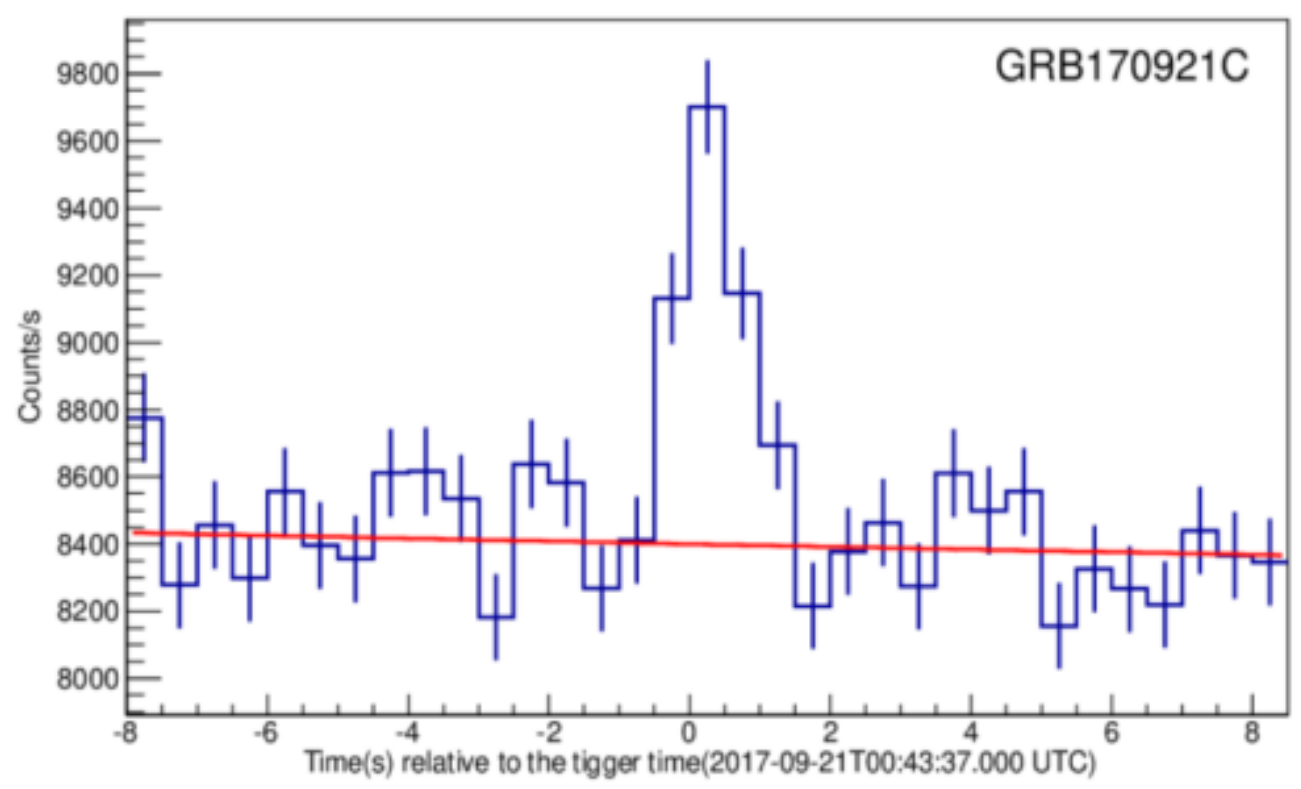

Figure 40 The short Gamma-ray Burst GRB 170921C detected by *Insight*-HXMT with a significance of 12 σ .

## 10. SUMMARY

As China's first X-ray astronomical satellite, *Insight*-HXMT covers energy bands of 1-15, 5-30, 20-250 keV for pointed and scanning observations, and 200-3000 keV as an all-sky monitor. *Insight*-HXMT completed its performance verification and calibration phases during a time period from June 15 to November 15 2017, and entered into normal observation phase by the end of 2017. Until June 2018, it has carried out more than 20 Galactic plane scan survey and monitoring, performed 183 normal and 13 ToO observations on 44 bright sources. More than 60 GRBs have been detected and the GW EM follow-up observations put stringent constraints to soft gamma-ray emissions from the counterpart of GW170817. Some early science results have proved that the performance of the *Insight*-HXMT satellite with its instruments and science operation with its user support software meets the mission requirements.

## ACKNOWLEDGEMENTS

This work made use of the data from the *Insight*-HXMT mission, a project funded by China National Space Administration (CNSA) and the Chinese Academy of Sciences (CAS). The authors thank support from the National Key Research and Development Program of China (2016YFA0400800), the Strategic Priority Research Program of the Chinese Academy of Sciences (Grant No. XDA04010202, XDA04010300 and XDB23040400), and the Chinese NSFC U1838201 and U 1838102.

## REFERENCES

[1]: Giacconi, R., Kellogg, E., Gorenstein, P., Gursky, H., and Tananbaum, H. 1971, ApJ, 165, L27
[2]: Bradt, Ohashi, and Pounds 1992, Ann. Rev. of A & A, 30, 391

[3]: IBIS: The Imager on-board INTEGRAL, Ubertini, P.; Lebrun, F.; Di Cocco, G.; Bazzano, A.; Bird, A. J.; Broenstad, K.; Goldwurm, A.; La Rosa, G.; Labanti, C.; Laurent, P.; Mirabel, I. F.; Quadrini, E. M.; Ramsey, B.; Reglero, V.; Sabau, L.; Sacco, B.; Staubert, R.; Vigroux, L.; Weisskopf, M. C.; Zdziarski, A. A., Astronomy and Astrophysics, v.411, p.L131-L139 (2003)
[4]: "The Burst Alert Telescope (BAT) on the Swift MIDEX Mission",
Barthelmy, S. D., Barbier, L. M., Cummings, J. R., Fenimore, E. E., Gerhels, N., Hullinger, D., Krimm, H. A., Markqardt, C. B., Palmer, D. M., Parsons, A., Goro, S., Suzuki, M., Takahashi, T., Tashiro, M., & Tueller, J, 2005, Space Sci. Rev., 120, 143-164
[5]: The Nuclear Spectroscopic Telescope Array (NuSTAR) High-energy X-Ray Mission
Harrison, Fiona A. et al.
The Astrophysical Journal, Volume 770, Issue 2, article id. 103, 19 pp. (2013).
[6]: C.J. Dai, M. Wu, Y.Q. Ma, Z.G. Lu, et al., Ch A&A 11, 179 (1987)
[7] :Z.G. Lu, J.Z. Wang, Y.G. Li, and P.R. Shen, NIMPA 362, 551 (1995)
[8] :T.P. Li, M. Wu, Z.G. Lu, J.Z. Wang et al., Ap&SS 205, 381 (1993)
[9] :T.P. Li and M. Wu, Ap&SS 206, 91 (1993)
[10]: T.P. Li and M. Wu, Ap&SS 215, 213 (1994)
[11] :T.P. Li, M. Wu, Z.G. Lu, J.Z Wang et al., AcASn 35, 105 (1994)
[12] :T.P. Li ExA 6, 63 (1995)
[13]: Hard X-ray luminosity function and absorption distribution of nearby AGN: INTEGRAL all-sky survey, Sazonov, S.; Revnivtsev, M.; Krivonos, R.; Churazov, E.; Sunyaev, R., Astronomy and Astrophysics, Volume 462, Issue 1, January IV 2007, pp.57-66
[14]: Tueller, J. et al. 2008, ApJ, 681, 113
[15]: 本专辑HE文章
[16]：本专辑 ME 文章
[17]：本专辑LE文章
[18]: A low-latency pipeline for GRB light curve and spectrum using Fermi/GBM near real-time data, Zhao, Yi; Zhang, Bin-Bin; Xiong, Shao-Lin; Long, Xi; Zhang, Qiang; Song, Li-Ming; Sun, Jian-Chao; Wang, Yuan-Hao; Li, Han-Cheng; Bu, Qing-Cui; Feng, Min-Zi; Li, Zheng-Heng; Wen, Xing; Wu, Bo-Bing; Zhang, Lai-Yu; Zhang, Yong-Jie; Zhang, Shuang-Nan; Shao, Jian-Xiong, Research in Astronomy and Astrophysics, Volume 18, Issue 5, article id. 057 (2018).
[19]: F.J. Lu, T.P. Li, X.J. Sun, M. Wu, and C.G. Page, Astron. & Astrophys. Supple. 115, 395 (1996).
[20] :S. Zhang, T.P. Li, and M. Wu, Astron. & Astrophys. 340, 62 (1998)
[21] :J. Guan, S. Zhang, et al., Science China, in preparation (2019)
[22] :X. Zhou, X.Q. Li, Y.N. Xie, C.Z. Liu, et al., Experimental Astronomy 38, 433 (2014)
[23]:S. Zhang, Y.P. Chen, Y.N. Xie, X.Q. Li, et al., Proc. SPIE 9144, 55 (2014)
[24]: Insight-HXMT science operations, Jia, S. M.; Ma, X.; Huang, Y.; Zhang, W. Z.; Ou, G.; Song, L. M.; Qu, J. L.; Zhang, S.; Chen, L., Proceedings of the SPIE, Volume 10704, id. 107041C 6 pp. (2018).
[25]: In-orbit calibration status of the Insight-HXMT, Li, Xiaobo; Song, Liming; Li, Xufang; Tan, Ying; Yang, Yanji; Ge, Mingyu, Proceedings of the SPIE, Volume 10699, id. 1069969 11 pp. (2018).
[26]: Huang, Y.; Qu, J. L.; Zhang, S. N.; Bu, Q. C., et al, The Astrophysical Journal 868, 122 (2018)
[27]:B. P. Abbott et al., ApJL 848, 12 (2017)
[28]:T.P. Li, S.L. Xiong, S.N. Zhang, F.J. Lu, et al., Science China 61(3), 031011 (2018)
[29]: Insight-HXMT Observations of 4U 1636-536: Corona Cooling Revealed with Single Short Type-I X-Ray Burst, Chen, Y. P.; Zhang, S.; Qu, J. L.; Zhang, S. N.; et al., The Astrophysical Journal Letters, Volume 864, Issue 2, article id. L30, 5 pp. (2018).
[30]: Insight-HXMT observations of the Crab pulsar, Tuo, You-Li; Ge, Ming-Yu; Song, Li-Ming; Yan, Lin-Li; Bu, Qing-Cui; Qu, Jin-Lu, Research in Astronomy and Astrophysics, Volume 19, Issue 6, article id. 087 (2019).
[31]: Insight-HXMT Observations of Swift J0243.6+6124 during Its 2017–2018 Outburst, Zhang, Yue; Ge, MingYu; Song, LiMing; Zhang, ShuangNan; Qu, JinLu; et al. The Astrophysical Journal, Volume 879, Issue 1, article id. 61, 7 pp. (2019).